\documentclass[english,aps,superscriptaddress,prb,eqsecnum,notitlepage]{revtex4-1}
\usepackage{babel}
\usepackage{amsmath,amssymb}
\usepackage{leftidx}
\usepackage{graphicx}
\usepackage{color}
\usepackage{xcolor}
\usepackage{epsfig}
\usepackage{ulem}

\newcommand{\be}{\begin{equation}}
\newcommand{\ee}{\end{equation}}
\newcommand{\bea}{\begin{eqnarray}}
\newcommand{\eea}{\end{eqnarray}}
\newcommand*{\bb}{c}
\newcommand*{\ba}{\tilde{c}}
\newcommand{\app}{\frac{a}{p}}

\newcommand{\ket}[1]{\left| #1 \right>}

\newcommand{\abs}[1]{\left| #1 \right|}

\newcommand{\pda}{ {\phantom{\dagger}}}
\newcommand{\pst}{{\phantom{*}}}
\newcommand{\La}{{\Lambda_1}}
\newcommand{\Lb}{{\Lambda_2}}
\newcommand{\fO}{\mathcal{O}}        
\newcommand{\fQ}{\mathcal{Q}}
\newcommand{\oa}{\mathcal{A}}
\newcommand{\ob}{\mathcal{B}}
\renewcommand{\oc}{\mathcal{C}}

\begin{document}
\title{Self-conjugate representation  $\mbox{SU}(3)$  chains}
\author{Kyle Wamer}
\affiliation{Department of Physics and Astronomy and Stewart Blusson Quantum Matter Institute, University of British Columbia, 
Vancouver, B.C., Canada, V6T1Z1}
% \affiliation{Department of Physics and Astronomy and Quantum Matter Institute, University of British Columbia, 
%   Vancouver, B.C., Canada, V6T1Z1}
\author{Francisco H. Kim}
\affiliation{Institute of Physics, Ecole Polytechnique F\'ed\'erale de Lausanne (EPFL), CH-1015, Lausanne, Switzerland}
\author{Mikl\'os Lajk\'o}
\affiliation{Institute of Physics, Ecole Polytechnique F\'ed\'erale de Lausanne (EPFL), CH-1015, Lausanne, Switzerland}
\author{Fr\'ed\'eric Mila}
\affiliation{Institute of Physics, Ecole Polytechnique F\'ed\'erale de Lausanne (EPFL), CH-1015, Lausanne, Switzerland}
\author{Ian Affleck}
\affiliation{Department of Physics and Astronomy and Stewart Blusson Quantum Matter Institute, University of British Columbia, 
Vancouver, B.C., Canada, V6T1Z1}

\begin{abstract}
  It was recently argued that  $\mbox{SU}(3)$  chains in the $p$-box symmetric irreducible representation (irrep) exhibit a ``Haldane gap''
  when $p$ is a multiple of 3 and are otherwise gapless  [Nucl.\ Phys.\ B {\bf 924}, 508 (2017)]. We extend this argument to the self-conjugate
  irreps of  $\mbox{SU}(3)$  with $p$ columns of length 2 and $p$ columns of length 1 in the Young tableau  ($p=1$ corresponding to the adjoint irrep),
  arguing that they are always gapped but have spontaneously broken parity symmetry for $p$ odd but not even. 
\end{abstract}
\maketitle
\section{Introduction}
While SU(2) spin chains have been very extensively studied both theoretically and experimentally, higher
symmetry SU($n$) chains represent a new domain which may be experimentally accessible with cold atoms.\cite{WuPRL2003,HonerkampHofstetter2004,Cazalilla2009,gorshkov2010,BieriSU32012,Scazza2014,takahashi2012, Pagano2014,ZhangScience2014, CazalillaReyreview2014, Capponi_SUNreview_AnnPhys2016} 
Chains with spins in the fully symmetric p-box irreducible representation (irrep) of SU(3) were studied in Refs.\ \onlinecite{BykovNuclPhys2012, LajkoNuclPhys2017}. The Lieb-Schulz-Mattis-Affleck (LSMA)
theorem  \cite{LSM1961,AffleckLieb1986} implies that these models must either be gapless or have spontaneously broken translation symmetry
for $p\neq 3m$, with $m$ a positive integer. By mapping into a flag manifold $\sigma$-model at large $p$, with topological angles $\pm 2\pi p/3$, it was argued that for $p\neq 3m$,
the models renormalize to the $\mbox{SU}(3)_1$ Wess-Zumino-Witten (WZW)  conformal field theory, which was also verified by Monte Carlo calculations.\cite{LajkoNuclPhys2017} 
Important extensions of the field theory treatment were made in Refs.\ \onlinecite{Sulejmanpasic2018, OhmoriSeiberg2019}, consistent with the
same conclusion. Here we extend these arguments to the self-conjugate $\mbox{SU}(3)$  irreps. In this case,
the LSMA theorem fails, as the number of boxes in the Young tableaux is always divisible by 3. We again map the 
chains into a related flag manifold quantum field theory with topological terms at large $p$. Notably, the model is not Lorentz
invariant in this case due to unequal velocities for the Goldstone bosons which appear in the perturbative limit. The
topological angles are now $\pm p\pi$, equivalently $0$ for $p$ even and $\pm \pi$ for $p$ odd. We solve the field theory in the strong coupling limit,
obtaining a gapped phase with spontaneously broken parity symmetry for p odd but not even. 
We also present  AKLT type \cite{AKLT1988,GreiterRachel2007,*GreiterRachelSchuricht2007} ground states of generalized chain models which are gapped for all $p$
but exhibit spontaneously broken parity symmetry for $p$ odd but not even.\begin{footnote}{In the context of spin chains, parity symmetry is often referred to as mirror symmetry.}\end{footnote}

In Sec.~\ref{sec:lfwt} we present the ``flavour-wave theory'' calculations (analogous to Holstein-Primakoff spin wave theory for $\mbox{SU}(2)$). In
Sec.~\ref{sec:ft} we derive a non-Lorentz invariant flag manifold $\sigma$-model (NLI$\sigma$M) at large $p$. Its perturbative
spectrum agrees with the low energy sector of the flavour-wave theory spectrum, consisting of 6 Goldstone bosons with two
different velocities. We don't expect such Goldstone bosons to exist in the true
spectrum because the  $\mbox{SU}(3)$  symmetry should not be spontaneously broken in accordance with the Mermin-Wagner-Coleman theorem.\cite{MerminWagner1966, Coleman1973}
%
%\note{but we can have conserved  $\mbox{SU}(3)$  symmetry and gapless modes, no?}
 %
  In Sec.~\ref{sec:LSMA} we present the failure of the LSMA theorem for chains with these irreps.  
In Sec.~\ref{sec:strongcoupling} we solve the strong coupling limit of the field theory, obtaining
a gapped phase with spontaneously broken parity symmetry for topological angles $\pm \pi$ corresponding to $p$ odd. In Sec.~\ref{sec:MC} we present Monte Carlo results that show the absence of the $\mbox{SU}(3)_1$ critical point that was present in the case of fully symmetric chains. \cite{LajkoNuclPhys2017} 
% In Sec.~\ref{sec:phasediag} we propose a phase diagram of the underlying NRSM.
In Sec.~\ref{sec:AKLT} we propose  AKLT  states consistent with these conclusions, with the spontaneously broken symmetry, for $p$ odd, again being parity. Sec.~\ref{sec:conclusions} contains conclusions. We also provide several appendices including detailed calculations and possible ways  to further verify our findings. 

%%%%%%%%%%%%%% SECTION: LFWT
%%%%%%%%%%%%%
%%%%%%%%%%%%%
%%%%%%%%%%%%%

\section{\label{sec:lfwt}Linear Flavour-Wave Theory}

The linear flavour-wave theory (LFWT),\cite{N1984281,papa1988,JoshiZhang1999} which is analogous to the $\mathrm{SU}(2)$ spin-wave theory, is a method that can
be applied to $\mathrm{SU}(n)$ models. The nomenclature originates from the $\mathrm{SU}(3)$ flavour symmetries of
elementary particles. It can be applied to an ordered state to obtain the low-energy spectrum of the model.

The aim in this section is to derive the velocities of the Goldstone modes which will serve as a check for the
field-theoretical approach in Sec.~\ref{sec:ft}. % We will first study the adjoint rep ($p=1$) using the multiboson
% approach~[\onlinecite{Penc,Judit,Kim2017}], which will allow us to understand the emergence of artificial flat modes that are an
% artifact of the method.
To this end, we will use the bosonic representation for $\mathrm{SU}(3)$ introduced by Mathur and
Sen \cite{MathurSen2001} to obtain the spectrum for any self-conjugate irrep represented by the Young tableaux $[p,p]$
with $p$ two-box columns and $p$ one-box columns.

\subsection{\label{sec:lfwt-ms}Bosonic representation of Mathur and Sen}
\begin{figure}[h]
  % \centerline{\includegraphics[width=0.99\linewidth]{fig1-young-tab.pdf} }
  \centerline{\includegraphics[width=6cm]{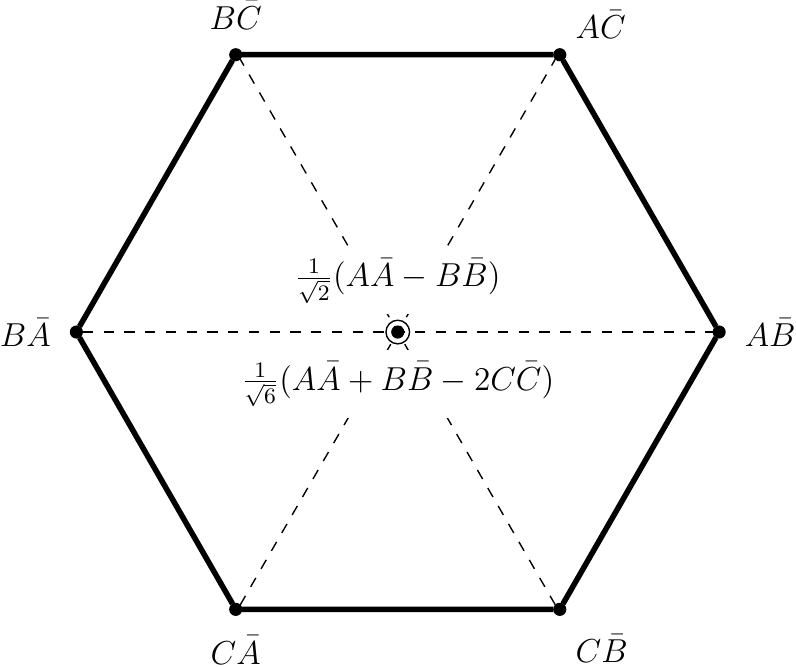} }
  \caption{The weight diagram of the adjoint irrep (p=1). The weight space in the middle is
    two-dimensional, and its two basis states can be chosen as in this figure. Similarly, the weight space on the second outer
    hexagon is always two-dimensional for any $p$ [\onlinecite{Pfeifer2003su3}].}
    \label{fig:weight-diag}
\end{figure}

 Following Mathur and Sen~\cite{MathurSen2001} , we write the spin operators in terms of two 3-component
commuting boson operators $a_\alpha$ and $b^\alpha$:
\begin{equation}
  \label{eq:su3-HW-S}
  \hat{S} ^{\alpha}_{\beta} = a ^{\alpha \dagger}_{} a ^{}_{\beta} - b ^{\dagger}_{\beta} b ^{\alpha}.
\end{equation}
The operators $a_{1}, a_{2}, a_{3}$ are related to the fundamental irrep $\mathbf{3}$ of $\mathrm{SU}(3)$ whose states
will be denoted by the flavours $A, B, C$, whereas the operators $b^{1}, b^{2}, b^{3}$ belong to its conjugate irrep
$\mathbf{\bar{3}}$ whose states will be labelled with $\bar{A}, \bar{B}, \bar{C}$. This construction naturally satisfies the $\mathrm{SU}(3)$ commutation relations
\begin{equation}
  \label{eq:su3-8-comm-rel}
    \left[ \hat{S} ^{\alpha}_{\beta}, \hat{S} ^{\mu}_{\nu} \right] = \delta ^{\mu}_{\beta} \hat{S} ^{\alpha}_{\nu} - \delta ^{\alpha}_{\nu}
    \hat{S} ^{\mu}_{\beta}.
\end{equation}
The $[p,p]$ self-conjugate irrep corresponds to states with p bosons of type $a$ and p bosons of type $b$, 
\begin{equation}
  \label{eq:su3-hw-constraints}
 \hat N_a = \sum\limits_{\alpha=1}^{3} a ^{\alpha\dagger}_{} a ^{}_{\alpha} = p, \quad \hat N_b= \sum\limits_{\alpha=1}^{3} b ^{\dagger}_{\alpha} b ^{\alpha}_{} = p,\quad  \sum_{\alpha=1}^3\hat{S}^\alpha_\alpha = \hat{N}_a-\hat{N}_b=0,
\end{equation}
but not all such states belong to the $[p,p]$ irrep. Take  the case of $p=1$ as an example. The $[1,1]$ irrep is 8 dimensional, but the states with one $a$ boson and one $b$ boson span a 9 dimensional subspace. The  states corresponding the self-conjugate irrep are shown in the weight diagram in Fig.~\ref{fig:weight-diag}, while the ninth state is $\ket{ A \bar A}+\ket{ B \bar B}+\ket{C \bar C}=(a^{1\dagger}b_1^\dagger+a^{2\dagger}b_2^\dagger+a^{3\dagger}b_3^\dagger )\ket{0}$, which actually belongs to the singlet irrep. In general the subspace spanned by states with p $a$ bosons and  p $b$ bosons  is a combination of all the self-conjugate irreps $[p',p']$ with $p'\leq p$ and the singlet irrep. To select the subspace corresponding to the $[p,p]$ irrep itself, we prove in Appendix \ref{appendixA} the following condition for any $\ket{\Psi}$ of p $a$ bosons and p $b$ bosons lying in $[p,p]$:
 \begin{equation} 
\begin{split}
\sum_\gamma a_\gamma b^\gamma \ket{\Psi}=0.
\end{split}
\label{eq:tracelessness}
\end{equation}
%We will inspect the final results of the flavour wave calculations with respect to this condition. 
  
In the following we will apply the LFWT to the $\mathrm{SU}(3)$ antiferromagnetic Heisenberg chain:
\begin{equation}
  \begin{aligned}
    \label{eq:su3-hw-H}
    \mathcal{H} =& J \sum\limits_{i} \sum\limits_{\alpha,\beta=1}^{3} \hat{S} ^{\alpha}_{\beta}(i) \hat{S}
    ^{\beta}_{\alpha}(i+1).
  \end{aligned}
\end{equation}

\subsection{Classical ground state}
\label{sec:lfwt-gs}

Mathur and Sen \cite{MathurSen2001} introduced spin-coherent states for the $[p,p]$ irrep using the $a$, $b$ bosons as 
\begin{equation}
\begin{split}
 |\vec z,\vec w\rangle := [(\vec z\cdot \vec a^\dagger )(\vec w\cdot \vec b^\dagger )]^p|0\rangle 
\end{split}
\end{equation}
where  $\vec z\cdot \vec w=0$ and $|\vec z|^2=|\vec w|^2=1$.  The first condition guarantees that these states satisfy the traceless condition of Eq.~\eqref{eq:tracelessness} and are thus  
in the correct irreducible  representation (see appendix \ref{appendixA}). The second condition is for normalization. These coherent states form an overcomplete set over the $[p,p]$ irrep. 
The expectation value of the spin operators reads as
\begin{equation}
\begin{split}
 \langle \vec z,\vec w|S^\alpha_\beta |\vec z,\vec w \rangle = p[z^{\alpha *}z_\beta^\pst -w^\alpha w_\beta^*].
\end{split}
\label{eq:coherentstate}
\end{equation}
The classical limit, which corresponds to the expectation of the quantum Hamiltonian in a direct product of spin-coherent states reads as
\begin{equation}
\begin{split}
 H
 &=Jp^2\sum_i\big[z^{\alpha *}_i z_{\beta ,i}^\pst-w^\alpha_i w_{\beta ,i}^*][z^{\beta *}_{i+1}z_{\alpha ,i+1}^\pst-
w^\beta_{i+1} w_{\alpha ,i+1}^*\big]\\
&=Jp^2\sum_i\big[|\vec z^*_i\cdot \vec z_{i+1}^\pst|^2+|\vec w^*_i\cdot \vec w_{i+1}^\pst|^2
-|\vec z_i^\pst\cdot \vec w_{i+1}^\pst|^2-|\vec w_i^\pst\cdot \vec z_{i+1}^\pst|^2\big].
\end{split}
\label{H}
\end{equation}

The classical groundstates are the two-sublattice states with
\be \vec z_{2n}^\pst=\vec w_{2n+1}^*=\vec \Phi^1,\ \  \vec w_{2n}^*=\vec z_{2n+1}=\vec \Phi^2\ee
where $|\vec \Phi^i|^2=1$ and $\vec \Phi^{1*}\cdot \vec \Phi^2=0$, giving an energy $-2Jp^2L$ where $L$ is 
the number of links. We can choose any of these states as  a starting point for the flavour-wave calculations, because they are all equivalent up to global  $\mbox{SU}(3)$  rotations. 

\begin{figure}[t]
  % \centerline{\includegraphics[width=0.99\linewidth]{fig1-young-tab.pdf} }
  \centerline{\includegraphics[width=8cm]{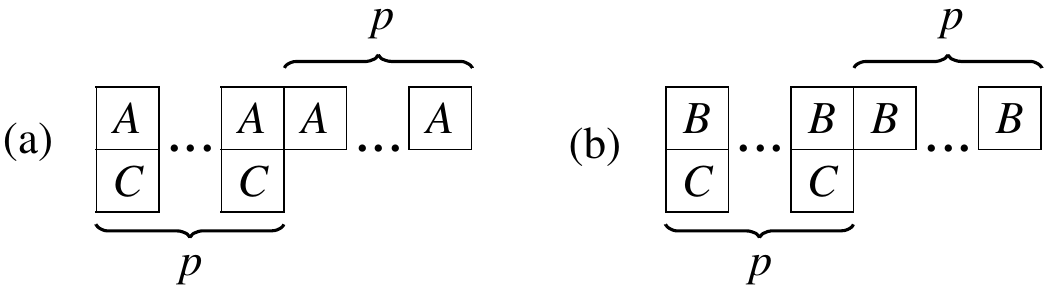} }
  \caption{\label{fig:lfwt-young} The two-sublattice classical ground state of the nearest neighbour Heisenberg model of Eq. \ref{eq:su3-hw-H}: (a) The state of sublattice $\La$ with $p$
    times $\bar{B}$ and $p$ times $A$. (b) The state of sublattice $\Lb$ with $p$ times $\bar{A}$ and $p$ times
    $B$.}
\end{figure}

\subsection{Flavour-wave spectrum}
\label{sec:lfwt-spectrum}
According to the above discussion, we  choose the classical ground state that is given by $p$
times $A$ and $p$ times $\bar{B}$ (or $(a^{1\dagger} b_2^\dagger)^p\ket{0}$ in bosonic language) on the sublattice $\Lambda_{1}$, and $p$ times $B$ and $p$ times $\bar{A}$ (or $(a^{2\dagger} b_1^\dagger)^p\ket{0}$)  on the
other sublattice $\Lambda_{2}$ in a N\'eel configuration. These two states are depicted in terms of the Weyl tableaux in
Fig.~\ref{fig:lfwt-young}. Using the notation of the coherent states these correspond to $\vec z_{2n}=\vec w_{2n+1}=(1,0,0)$ and $\vec w_{2n}=\vec z_{2n+1}=(0,1,0)$.

We now take the semi-classical limit by letting $p \rightarrow \infty$, just as in the
spin-wave calculations in which $S \rightarrow \infty$. Under this assumption of a large condensate of $A$, $\bar{B}$ on
$i \in \Lambda_{1}$ and $B$, $\bar{A}$ on $j \in \Lambda_{2}$, the constraints in Eq.~\eqref{eq:su3-hw-constraints} can be rewritten as
\begin{equation}
  \label{eq:su3-hw-constraints-HP}
  \begin{aligned}
    a ^{1\dagger}_{} (i) a ^{}_{1} (i) = & \,\,p  - [a ^{2\dagger}_{} (i) a ^{}_{2} (i) + a ^{3\dagger}_{} (i) a ^{}_{3} (i)],\\
    b ^{\dagger}_{2} (i) b ^{2}_{} (i) = & \,\,p  - [b ^{\dagger}_{1} (i) b ^{1}_{} (i) + b ^{\dagger}_{3} (i) b ^{3}_{} (i)],\\
    a ^{2\dagger}_{} (j) a ^{}_{2} (j) = & \,\,p  - [a ^{1\dagger}_{} (j) a ^{}_{1} (j) + a ^{3\dagger}_{} (j) a ^{}_{3} (j)],\\
    b ^{\dagger}_{1} (j) b ^{1}_{} (j) = & \,\,p  - [b ^{\dagger}_{2} (j) b ^{2}_{} (j) + b ^{\dagger}_{3} (j) b ^{3}_{} (j)],\\
  \end{aligned}
\end{equation}
% Let us define $\mu ^a_{\nu} := \sum\limits_{\mu=1}^{3} a ^{\dagger}_{\mu} a ^{}_{\mu} - a ^{\dagger}_{\nu} a ^{}_{\nu}$
% and $\mu ^b_{\nu} := \sum\limits_{\mu=1}^{3} b ^{\dagger}_{\mu} b ^{}_{\mu} - b ^{\dagger}_{\nu} b ^{}_{\nu}$ to
where we treat the bosons on the RHS as small fluctuations. %From here on, let us write the flavour indices as subscripts as we will shortly introduce superscript indices to indicate the sublattice of the bosons.

 The Holstein-Primakoff transformation reads as
\begin{equation}
  \label{eq:su3-hw-HP}
  \begin{aligned}
    a ^{1\dagger} (i), a ^{}_{1} (i) \quad \longrightarrow \quad & \sqrt{p - \sum\limits_{\alpha \neq 1}a
      ^{\alpha\dagger}(i) a ^{}_{\alpha}(i)},\\
    b ^{\dagger}_{2} (i), b^{2} (i) \quad \longrightarrow \quad & \sqrt{p - \sum\limits_{\alpha \neq
        2}b^{\dagger}_{\alpha}(i) b ^{\alpha}(i)},\\
    a ^{2\dagger} (j), a ^{}_{2} (j) \quad \longrightarrow \quad & \sqrt{p - \sum\limits_{\alpha \neq
        2}a^{\alpha\dagger}_{}(j) a ^{}_{\alpha}(j)},\\
    b ^{\dagger}_{1} (j), b ^{1} (j) \quad \longrightarrow \quad & \sqrt{p - \sum\limits_{\alpha \neq
        1}b^{\dagger}_{\alpha}(j) b ^{\alpha}(j)}.\\
  \end{aligned}
\end{equation}
We can now apply this transformation on the Hamiltonian~\eqref{eq:su3-hw-H} written with the boson
operators~\eqref{eq:su3-HW-S}, which gives the quadratic Hamiltonian at the order $\mathcal{O}(p)$:
\begin{equation}
  \label{eq:su3-hw-H2}
  \begin{aligned}
    \mathcal{H}^{(2)} = J p \sum\limits_{i \in \Lambda_{1}}\sum\limits_{\substack{j \in \Lb\\ j=i\pm1}} &\left\{ \Big[ 2 a ^{2\dagger}(i) a ^{}_{2}(i) + 2 a ^{1\dagger}(j) a
        ^{}_{1}(j) + 2 b ^{\dagger}_{1}(i) b ^{1}(i) + 2 b ^{\dagger}_2(j) b ^{2}(j) \right. \\
    &\quad +  a ^{2\dagger}(i) a
    ^{1\dagger}(j) -  a ^{2\dagger}(i) b ^{\dagger}_{2}(j) -  b ^{\dagger}_{1}(i) a ^{1\dagger}(j) +  b ^{\dagger}_{1}(i) b ^{\dagger}_{2}(j)\\
    &\quad  + a
      ^{}_{2}(i) a ^{}_{1}(j) - a ^{}_{2}(i) b ^{2}(j) - b ^{1}(i) a ^{}_{1}(j) + b ^{1}(i) b ^{2}(j) \Big] \\
    &+ \left[ a ^{3\dagger}(j) a ^{}_{3}(j) + b ^{\dagger}_{3}(i) b ^{3}(i) - b ^{\dagger}_{3}(i) a
      ^{3 \dagger}(j) - b ^3(i) a ^{}_{3}(j) \right]\\
    &+ \left. \left[ a ^{3\dagger}(i) a ^{}_{3}(i) + b ^{\dagger}_{3}(j) b ^{3}(j) - a ^{3 \dagger}(i) b
        ^{3 \dagger}(j) - a ^{}_{3}(i) b ^{3}(j) \right] \right\}.
  \end{aligned}
\end{equation}
We now use the Fourier transform,
\begin{equation}
  \label{eq:su3-hw-FT}
  a^{}_{\alpha}(l) = \sqrt{\frac{2}{L}} \sum\limits_{k\in \text{RBZ}}
  a_{\alpha}(k, \Lambda_l)e^{-i k r_{l}}, 
  \qquad b^{\beta}(l) = \sqrt{\frac{2}{L}}
  \sum\limits_{k\in \text{RBZ}} b^{\beta}(k,\Lambda_l)e^{-i k r_{l}},\\
\end{equation}
where $k$ runs over the reduced Brillouin zone (RBZ), $L$ is the total number of sites and
$\Lambda_l \in \left\{\La, \Lb \right\}$ is the sublattice index keeping track of the sublattice  of site $l$. The
quadratic Hamiltonian~\eqref{eq:su3-hw-H2} is then given by (the left superscript  stands for transpose)
\begin{equation}
  \label{eq:su3-hw-bog-H2-nondiag}
  \mathcal{H}^{(2)} = J p \sum\limits_{k\in \text{RBZ}} \sum\limits_{\alpha=1}^{3} \left(
    \mathbf{\bb}^{\alpha\dagger}_{k},\mathbf{\bb}^{\alpha}_{-k} \right) M^{\alpha}_{k}
  \begin{pmatrix}
    \ltrans{\mathbf{\bb}^{\alpha}_{k}}\\
    \ltrans{\mathbf{\bb}^{\alpha\dagger}_{-k}}
  \end{pmatrix}
\end{equation}
where
\begin{equation}
  \begin{aligned}[c]
    \mathbf{\bb}^{1\dagger}_{k} :=& \Bigl(a^{2,\dagger}(k, \La), b^{ \dagger}_{1}(k, \La), a^{1\dagger}(k, \Lb), b^{\dagger}_{2}(k,\Lb) \Bigr),\\
    \mathbf{\bb}^{2\dagger}_{k} :=& \Bigl(a^{3\dagger}(k, \La ), b^{\dagger}_{3}(k,\Lb) \Bigr),\\
    \mathbf{\bb}^{3\dagger}_{k} :=& \Bigl(a^{3 \dagger}(k, \Lb), b^{\dagger}_{3}(k,\La) \Bigr),\\
  \end{aligned} \qquad
  \begin{aligned}[c]
    \mathbf{\bb}^{1}_{-k} :=& \Bigl(a_{2}(-k, \La), b^{1}(-k,\La), a_{1}(-k,\Lb), b^{2}(-k,\Lb) \Bigr),\\
    \mathbf{\bb}^{2}_{-k} :=& \Bigl(a_{3}(-k, \La), b^{3}(-k, \Lb) \Bigr),\\
    \mathbf{\bb}^{3}_{-k} :=& \Bigl(a_{3}(-k,\Lb), b^{3}(-k,\La) \Bigr),
  \end{aligned}
\end{equation}
and
\begin{equation}
  \begin{aligned}
    M^{1}_{k} :=& 
    \begin{pmatrix}
      A^{1} & B^{1}_{k}\\
      B^{1\dagger}_{k} & A^{1}
    \end{pmatrix}, \qquad
    % A^{1}_{k} := 2p \mathbbm{1}_{4} , \qquad
    A^{1} := 2\, \mathbf{1}_{4}, \qquad
    B^{1}_{k} :=
    \begin{pmatrix}
      0 & 0 & \gamma_{k} & -\gamma_{k} \\
      0 & 0 & -\gamma_{k} & \gamma_{k}  \\
      \gamma_{k} & -\gamma_{k} & 0 & 0 \\
      -\gamma_{k} & \gamma_{k} & 0 & 0 \\
    \end{pmatrix},\\
    M^{3}_k := M^{2}_{k} :=& 
    \begin{pmatrix}
      A^{2} & B^{2}_{k}\\
      B^{2}_{k} & A^{2}
    \end{pmatrix}, \qquad \
    A^{2} :=  \mathbf{1}_{2}, \qquad \ \
    % \begin{pmatrix}
    %   p & 0 \\
    %   0 & p \\
    % \end{pmatrix}, \qquad
    B^{2}_{k} :=
    \begin{pmatrix}
      0 & - \gamma_{k} \\
      - \gamma_{k} & 0 \\
    \end{pmatrix},\\
  \end{aligned}
\end{equation}
where the geometrical factor $\gamma_{k} := \cos (ka)$ has been introduced. We now diagonalize the system by using the generalized Bogoliubov transformation.\cite{Ripka1986} Then the positive eigenvalues
of the matrices
\begin{equation}
  % M^{\alpha}_{\mathbf{k}}{}' :=& 
                                 \begin{pmatrix}
                                   A^{\alpha}_{} & B^{\alpha}_{k}\\
                                   -B^{\alpha \dagger}_{k} & -A^{\alpha}_{}
                                 \end{pmatrix}
\end{equation}
yield the frequencies $\omega_\mu$ of the system. Hence, we finally obtain the diagonalized Hamiltonian
\begin{equation}
  \label{eq:su3-hw-h2-diag}
    \begin{aligned}
      \mathcal{H}^{(2)} = J \sum\limits_{k\in \text{RBZ}} &\left\{ \sum\limits_{\mu=1}^{8} \omega_{\mu}(k) \left(
          \ba^{\dagger}_{\mu}(k) \ba^{}_{\mu}(k) + \frac{1}{2} \right) \right\} + \text{const.}
    \end{aligned}
\end{equation}
where the bosons $\ba^{}_{\mu}$ are the new Bogoliubov bosons, and
\begin{equation}
  \label{eq:su3-hw-w-1D-self-conjugated}
    \begin{aligned}[r]
      \omega_{1,2}(k) &= 4 p \abs{\sin (ka)}, \quad \omega_{3,4}(k) = 2 p \abs{\sin (ka)},\\
      \omega_{5,6}(k) &= 2 p \abs{\sin (ka)}, \quad \omega_{7,8}(k) = 4p,\\
    \end{aligned}
\end{equation}
yielding 2 different types of Goldstone modes and 2 flat modes. In Appendix \ref{app:modes} we give a detailed explanation of both the dispersive and the flat modes, and their relations to the spin generators. Hence, we finally observe that the dispersion relations related to the Goldstone modes are $\omega_{1,\dots,6}$. The
velocities of these six Goldstone modes are given by
\begin{equation}
  \label{eq:su3-hw-v}
v_1:=v_2:=4apJ, \quad v_3:=v_4:=v_5:=v_6:=2apJ.
\end{equation}
While the Goldstone modes are absent in the actual spectrum, the true low energy excitations will have a mass that is exponentially suppressed in $p$, and still well below the $\fO(p)$ energy scale of the flat modes. Indeed, this was shown to be true in the case of equal velocities in Ref.~\onlinecite{LajkoNuclPhys2017}, using the coupling constant beta function. Since unequal velocities will only modify the beta function by a $p$-independent function of their ratios, we expect the same exponential dependence to hold in the present theory. %{\color{red} is this okay? -- if I complete the RG for unequal velocities soon, we can add the explicit mass gap here.}

%\subsection{\note{Further explanation for internal use, analogy with the spin one chain}}
%
%Consider the nearest neighbour $S=1$ antiferromagnetic Heisenberg chain. The classical ground state is the $\ket{1\bar11\bar1\dots}$ N\'eel-ordered state. We can do spin wave calculations in two different ways. One using the $b_{\uparrow(\downarrow)}$ Schwinger bosons, the other way is to introduce a boson for each spin state$ (c_1^{(\dagger)}, c_0^{(\dagger)},c^{(\dagger)}_\oneb)$. 
%In the Sch operators in the two representations will read as
%
%\begin{equation}
%\begin{split}
%S^z =   \frac{1}{2} (b^\dagger_\uparrow b_\uparrow^\pda -b_\downarrow^\dagger b^\pda_\downarrow),  \quad
%S^+=  \sqrt{2}  (b^\dagger_\uparrow b_\downarrow^\pda), \quad
%S^-= \sqrt{2}  (b^\dagger_\downarrow b_\uparrow^\pda).
%\end{split}
%\end{equation}

%%%%%%%%%%%%%
%%%%%%%%%%%%%
%%%%%%%%%%%%%

\section{\label{sec:ft}Mapping to Field Theory}
%\subsection{\label{sec:ft-coherent}Notation and Coherent States}
Using the coherent states introduced earlier we can carry out a spin-coherent state path integral approach  \cite{KlauderPRD1979,GnutzmannKus1998, ShibataTakagi1999,LajkoNuclPhys2017} on the quantum spin Hamiltonian of Eq.~\eqref{eq:su3-hw-H} 
The imaginary Berry's phase term in the action will be
\be S_{B,n}=-p\int d\tau \left[ \vec z_n^* \cdot {d\over d\tau}\vec z_n^\pst+\vec w_n^*\cdot{d\over d\tau} \vec w_n^\pst\right] ,\ee
while the Hamiltonian part becomes 
\begin{equation*}
\begin{split}
 H
% &=Jp^2\sum_i\big[z^{\alpha *}_i z_{\beta ,i}^\pst-w^\alpha_i w_{\beta ,i}^*][z^{\beta *}_{i+1}z_{\alpha ,i+1}^\pst-
%w^\beta_{i+1} w_{\alpha ,i+1}^*\big]\\
&=Jp^2\sum_n\big[|\vec z^*_n\cdot \vec z_{n+1}^\pst|^2+|\vec w^*_n\cdot \vec w_{n+1}^\pst|^2
-|\vec z_n^\pst\cdot \vec w_{n+1}^\pst|^2-|\vec w_n^\pst\cdot \vec z_{n+1}^\pst|^2\big],
\end{split}
\end{equation*}
as discussed in Sec.~\ref{sec:lfwt-gs}. In this approach we parametrize fluctuations around the classical ground state manifold 
\be \vec z_{2j}^\pst=\vec w_{2j+1}^*=\vec \Phi^1,\ \  \vec w_{2j}^*=\vec z_{2j+1}^\pst=\vec \Phi^2,\label{phi}\ee
where $|\vec \Phi^i|^2=1$ and $\vec \Phi^{1*}\cdot \vec \Phi^2=0$. In the classical ground state
%\be \vec z_{2n+1}(\tau )\approx \vec w_{2n}^*(\tau ),\ \  \vec w_{2n+1}(\tau )\approx \vec z_{2n}^*(\tau ),\ee
%then
\be S_{B,2j}+S_{B,2j+1}\approx -p\int d\tau \left[\vec z^*_{2j}\cdot {d\over d\tau }\vec z_{2j}^\pst
+\vec w^*_{2j}\cdot {d\over d\tau } \vec w_{2j}^\pst+\vec w_{2j}^\pst\cdot {d\over d\tau }\vec w_{2j}^*
+\vec z_{2j}^\pst\cdot {d\over d\tau } \vec z_{2j}^*
\right]=-p\int d\tau {d\over d\tau}[|\vec z_{2j}^\pst|^2+|\vec w_{2j}^\pst|^2]=0. \ee

Now, if we allow the $\vec \Phi^1$ and $\vec \Phi^2$ fields to change from site to site,
\be \vec z_{2j}^{\phantom{1}}=\vec \Phi^1_{2j},\ \  \vec w_{2j}^{\phantom{1}}=\vec \Phi^{2*}_{2j},\ \  \vec z_{2j+1}^{\phantom{1}}=\vec \Phi^2_{2j+1},\ \  
\vec w_{2j+1}^{\phantom{1}}=\vec \Phi^{1*}_{2j+1},\label{pdef}\ee
the Hamiltonian becomes
\be H=Jp^2\sum_n[|\vec \Phi^{1*}_n\cdot \vec \Phi^2_{n+1}|^2+|\vec \Phi^{2*}_n\cdot \vec \Phi^1_{n+1}|^2-|\vec \Phi^{1*}_n\cdot \vec \Phi^1_{n+1}|^2-|\vec \Phi^{2*}_n\cdot \vec \Phi^2_{n+1}|^2].
\ee
$\vec \Phi^{i*}_n\cdot \vec \Phi^j_n=\delta^{ij}$  must be strictly enforced on every site since it follows from the condition that we are in the correct irrep.  We thus may combine the $\vec \Phi^i$ on each site to 
define a unitary matrix:
\be W_n := \left(\begin{array}{c} \vec \Phi^1_n \\ \vec \Phi^2_n \\ \vec \Phi^3_n \end{array} \right) \ee
where 
\be \vec \Phi^3_n := \vec \Phi^{1*}_n\times \vec \Phi^{2*}_n\ee
is uniquely defined from $\vec \Phi^1$ and $\vec \Phi^2$ in order that $W_n$ be an  $\mbox{SU}(3)$ matrix. Note that this is unlike the path integral approach for $\mbox{SU}(3)$  chains in the fully symmetric irrep,\cite{LajkoNuclPhys2017} where  $\vec \Phi^1$, $\vec \Phi^2$ and $\vec \Phi^3$ were defined on 3 neighbouring sites so they did not have to be all mutually exactly orthogonal.

 Using the $W_n$ matrices the Hamiltonian term can be written as
\be H=-Jp^2\sum_n \hbox{tr} (W^\dagger_n\Lambda W_n^\pda W^\dagger_{n+1}\Lambda W_{n+1}^\pda),\ee
where 
\be \Lambda := \left( \begin{array}{rrr} 1&0&0\\ 0&-1&0\\ 0&0&0\end{array}\right).\ee
The Berry's phase term in the action becomes
\be S_B=-p\sum_n(-1)^n\int d\tau [\vec \Phi^{1*}_n\cdot \partial_\tau \vec \Phi^1_n-\vec \Phi^{2*}_n\cdot \partial_\tau \vec \Phi^2_n].\label{ptt}\ee
Assuming the $\vec \Phi^i_n$ vary smoothly,  the classical Hamiltonian density becomes
\be a{\cal H}_{cl}/(Jp^2)=a^2|\vec \Phi^{1*}\cdot \partial_x\vec  \Phi^2|^2+a^2|\vec \Phi^{2*}\cdot \partial_x\vec  \Phi^1|^2-|1+\vec \Phi^{1*}\cdot [a\partial_x\vec \Phi^1+(1/2)a^2\partial_x^2\vec \Phi^1]|^2
-|1+\vec \Phi^{2*}\cdot [a\partial_x\vec \Phi^2+(1/2)a^2\partial_x^2\vec \Phi^2]|^2.
\ee
The single derivative terms cancel, and after simple tranformations using the orthogonality of $\vec \Phi_1$ and $\vec \Phi_2$, this becomes
\be {\cal H}_{cl}/(Jp^2a)=|\partial_x\vec \Phi^1|^2-|\vec \Phi^{1*}\cdot \partial_x\vec \Phi^1|^2+|\partial_x\vec \Phi^2|^2-|\vec \Phi^{2*}\cdot \partial_x\vec \Phi^2|^2
+2|\vec \Phi^{1*}\cdot \partial_x\vec  \Phi^2|^2.\label{2sd}
\ee
All terms can be seen to be  invariant under the $\vec \Phi^{i}(x,\tau) \to \vec \Phi^{i}(x,\tau) e^{i \vartheta_{i}(x,\tau)}$ gauge transformation.\cite{LajkoNuclPhys2017} 
$\mathcal{H}_{cl}$ can be rewritten with purely off-diagonal terms using the fact that the $\vec \Phi^{i}$ on a given site form an orthonormal basis:
  \begin{equation}
\begin{split}
 \left( \left| \partial_\mu \vec{\Phi}^{i\phantom{*}}\right|^2 -  \left| \vec{\Phi}^{i*} \cdot \partial_\mu \vec{\Phi}^{i\phantom{*}} \right|^2 \right)=  \left|  \vec{\Phi}^{i+1*} \cdot \partial_\mu \vec{\Phi}^{i\phantom{*}} \right|^2+  \left|  \vec{\Phi}^{i-1*} \cdot \partial_\mu \Phi^{i\phantom{*}} \right|^2.
\end{split}
\end{equation} 
Therefore
\be {\cal H}_{cl}/(Jp^2a)= 4|\vec \Phi^{1*}\cdot \partial_x\vec  \Phi^2|^2+|\vec \Phi^{1*}\cdot \partial_x\vec  \Phi^3|^2+|\vec \Phi^{2*}\cdot \partial_x\vec  \Phi^3|^2.\label{Hcl}\ee
In terms of the $W$ matrix this is
\be {\cal H}_{c}/(Jp^2a)=\hbox{tr}[4\Lambda_2\partial_xWW^\dagger \Lambda_1W\partial_xW^\dagger 
+\Lambda_3\partial_xWW^\dagger \Lambda_1W\partial_xW^\dagger +\Lambda_3\partial_xWW^\dagger \Lambda_2W\partial_xW^\dagger ],\ee
where
\be
\Lambda_1 := \left(\begin{array}{ccc} 1&0&0\\ 0&0&0\\ 0&0&0\end{array}\right),\ \
  \Lambda_2 := \left(\begin{array}{ccc} 0&0&0\\ 0&1&0\\ 0&0&0\end{array}\right),\ \
    \Lambda_3 := \left(\begin{array}{ccc} 0&0&0\\ 0&0&0\\ 0&0&1\end{array}\right).\label{defLambda}
  \ee

\subsection{Uniform and staggered parts}
We now give a parametrization of the $W_{2j}$ and $W_{2j+1}$ matrices. We follow concepts used in both the SU(2)  \cite{Sierra1997,Cabra2004} and in the fully symmetric  $\mbox{SU}(3)$  case  \cite{LajkoNuclPhys2017} as well. The $\mbox{SU}(3)$ spin operators correspond to the spin matrices in the path integral 
\be \mathcal{S}^\alpha_{\beta ,n}=p(-1)^n (W^\dagger_n\Lambda W_n^\pda)^\alpha_\beta =p(-1)^n(\Phi^{1\alpha*}_n\Phi^1_{\beta ,n}-\Phi^{2\alpha*}_n\Phi^2_{\beta ,n}).
\ee
The sign alternation arises from the different role of $\vec \Phi^1$ and $\vec \Phi^2$ on the two sublattices as shown in Eq.~\eqref{pdef}. Similarly to the SU(2) case, we introduce staggered and uniform parts of these spin matrices inside the two site unit cell.
The staggered part of the spin corresponds to the uniform, slowly changing part of the $\vec \Phi$ fields, while the uniform part of the spin matrices corresponds to staggered terms in the $\vec \Phi$ field. This needs to be done while maintaining strict orthogonality of the $\vec \Phi^i_n$ on each site $n$. 
We do this, using a two site basis, by writing
\be W_{2j}^\pda =V_j^\pda U_j^\pda ,\ \  W_{2j+1}^\pda =V_j^\dagger U_j^\pda, \ee
where $U_j$ and $V_j$ are both unitary matrices, defined for the unit cell $j$. We write
\be U_j=\left(\begin{array}{c} \vec \phi^1_j \\ \vec \phi^2_j \\ \vec \phi^3_j \end{array} \right),\label{Udef} \ee
which corresponds to the staggered part of the spins, describing a fully satisfied bond inside the unit cell. 
On the other hand, the unitary matrix $V_j$  describes fluctuations away from the perfect N\'eel order inside the unit cell, i.e.\ it corresponds to the uniform part of the spins. Since the presence of a uniform spin component gives a finite energy, we can take $V_j$  to be close to the identity, and thus the uniform part to be small compared to the staggered part.  We can write $V_j$ in terms of the Gell-Mann matrices, 
\be V=e^{i\vec \vartheta \cdot \vec T}.\ee
We drop the diagonal Gell-Mann matrices, $T_3$ and $T_8$, since these correspond to gauge transformations that leave the action invariant\footnote{This is justified thanks to the factorization formula of SU(3) matrices discussed in Eq.\ (5.8) and Appendix G of Ref.\ \onlinecite{LajkoNuclPhys2017} }. To leading approximation, we can expand the off-diagonal terms in $V$ to first 
order in the $\theta$'s and the diagonal terms to second order.  It is convenient to write
\be \vartheta_2+i\vartheta_1=\app L_{12},\ \  \vartheta_5+i\vartheta_4= \app L_{13},\ \  \vartheta_7+i\vartheta_6=\app L_{23}.\ee
Then we can approximate:
\begin{equation}
\begin{split} V&\approx \left(\begin{array}{ccc} 1-\frac{a^2}{2p^2}(|L_{12}|^2+|L_{13}|^2)&\app L_{12}&\app L_{13}\\
-\app L_{12}^*&1-\frac{a^2}{2p^2}(|L_{12}|^2+|L_{23}|^2)&\app L_{23}\\
-\app L_{13}^*&-\app L_{23}^*&1-\frac{a^2}{2p^2}(|L_{31}|^2+|L_{23}|^2)\end{array}\right)\\
V^\dagger &\approx \left(\begin{array}{ccc} 1-\frac{a^2}{2p^2}(|L_{12}|^2+|L_{13}|^2)&-\app L_{12}&-\app L_{13}\\
\app L_{12}^*&1-\frac{a^2}{2p^2}(|L_{12}|^2+|L_{23}|^2)&-\app L_{23}\\
\app L_{13}^*&\app L_{23}^*&1-\frac{a^2}{2p^2}(|L_{31}|^2+|L_{23}|^2)\end{array}\right).
\end{split}
\end{equation}
Next order corrections will be off-diagonal terms of $\fO[(a/p)^2]$ and diagonal terms of $\fO[(a/p)^3]$. Thus
\be
\def\arraystretch{1.5} \vec W_{2j/2j+1}=\left(\begin{array}{c}\vec \Phi^1_{2j/2j+1}\\
\vec \Phi^2_{2j/2j+1}\\
\vec \Phi^3_{2j/2j+1}\end{array}\right) 
\approx 
 \left(\begin{array}{c} 
\big(1-\frac{a^2}{2p^2}(|L_{12,j}|^2+|L_{13,j}|^2)\big)\vec \phi^1_j\\ 
\big(1-\frac{a^2}{2p^2}(|L_{12,j}|^2+|L_{23,j}|^2)\big)\vec \phi^2_j\\
\big(1-\frac{a^2}{2p^2}(|L_{13,j}|^2+|L_{23,j}|^2)\big)\vec \phi^3_j
\end{array}\right)
\pm {a\over p}\left(\begin{array}{c} L_{12,j}\vec \phi^2_j+L_{13,j}\vec \phi^3_j\\
-L_{12,j}^*\vec \phi^1_j+L_{23,j}\vec \phi^3_j\\
-L_{13,j}^*\vec \phi^1_j-L_{23,j}^*\vec \phi^2_j
\end{array}\right).
\ee

\subsection{Integrating out the $L$ variables}

With the above parametrization the total action reads as (the detailed calculations are provided in Appendix \ref{app:SLU})

\begin{equation}
\begin{split}
S=&{} \frac{1}{2a}\int d\tau dx \Bigg[32a^2 J |L_{12}|^2 +8a^2 J |L_{23}|^2 +8a^2J |L_{13}|^2 
+8a^2 p^2 J| \vec \phi^{2^*}\cdot\partial_x \vec \phi^1|^2 \\
&+4a^2p^2 J\big(|\partial_x\vec \phi^1|^2-|\vec \phi^{1*}\cdot \partial_x \vec \phi^1|^2 \big)
+4a^2p^2J \big(|\partial_x \vec \phi^2|^2-|\vec \phi^{2*}\cdot \partial_x \vec \phi^2|^2 \big)  \\
&+ a p L_{12} \left( 16 a J(\vec \phi^2 \cdot\partial_x \vec \phi^{1*})+
\frac{4}{p} (\vec \phi^{1*} \cdot \partial_\tau \vec \phi^2) \right) + a p L_{12}^* 
\left( 16 aJ (\vec \phi^{2*}\cdot \partial_x \vec \phi^1)+\frac{4}{p} (\vec \phi^{2*} \cdot \partial_\tau 
 \vec \phi^1) \right) \\
&-a p L_{23} \left( 4 a J(\vec \phi^{2*} \cdot\partial_x \vec \phi^3)+\frac{2}{p} 
(\vec \phi^{2*}\cdot \partial_\tau \vec \phi^3) \right) - a p L_{23}^* \left( 4 a J(\vec \phi^2 \cdot\partial_x \vec \phi^{3*})
+\frac{2}{p} (\vec \phi^{3*} \cdot \partial_\tau \vec \phi^2) \right) \\
&+a p L_{13}^* \left( 4 a J(\vec \phi^{3*} \cdot\partial_x \vec \phi^1)
+\frac{2}{p} (\vec \phi^{3*} \cdot \partial_\tau \vec \phi^1) \right) + a p L_{13} 
\left( 4 a J(\vec \phi^3 \cdot\partial_x \vec \phi^{1*})+\frac{2}{p} (\vec \phi^{1*} \cdot \partial_\tau \vec \phi^3) \right)
\Bigg].
\end{split}
\label{eq:SLU}
\end{equation}
Integrating out the $L_{ij}$ fields gives
\begin{equation}
\begin{split}
 S=\int dx d\tau \Bigg( &ap^2J\Big[ 4|\vec \phi^{2*}\cdot \partial_x\vec \phi^1|^2+|\vec \phi^{3*}\cdot \partial_x\vec \phi^2|^2
+|\vec \phi^{1*}\cdot \partial_x\vec \phi^3|^2\Big]  \\
&+{1\over4aJ}\Big[|\vec \phi^{2*}\cdot \partial_\tau \vec \phi^1|^2+|\vec \phi^{2*}\cdot \partial_\tau \vec \phi^3|^2
+|\vec \phi^{1*}\cdot \partial_\tau \vec \phi^3|^2\big]  \\
&+i\pi p[2q_{12}+q_{32}+q_{13}]\Bigg)
\label{eq:symmaction}
\end{split}
\end{equation}
where
\be q_{mn} := {1\over 2\pi i}\epsilon_{\mu \nu}\int dx d\tau (\vec \phi^m\cdot \partial_\mu \vec \phi^{n*})
(\vec \phi^{m*}\cdot \partial_\nu \vec \phi^{n})=-q_{nm}.
\label{eq:qmn}\ee
It follows from the completeness of the $\vec \phi^m$ vectors that 
\be Q_n=(q_{n,n-1}+q_{n,n+1})\ee
where 
\be Q_n := {1\over 2\pi i}\int dx d\tau \epsilon_{\mu \nu}\partial_\mu \vec \phi^{n}\cdot \partial_\nu \vec \phi^{n*}\ee
is an integer valued topological charge \cite{BergLuscher1981,UedaShannon2016} for the field $\vec \phi^n$ . Thus we write the imaginary part of the action as
\be S_{im}=i\pi p(Q_1-Q_2).\ee
Note that the space derivative terms of Eq.~\eqref{eq:symmaction} are identical to the classical Hamiltonian of Eq.~\eqref{Hcl} upon ignoring the 
difference between the $\vec \Phi^n$ and the $\vec \phi^n$.

In terms of the unitary matrix $U$, defined in Eq.~\eqref{Udef}, the real part of the Lagrangian density  can be written:
\begin{equation}
\begin{split}
 {\cal L}_{\hbox{real}}=&{}ap^2J\hbox{tr}\left[4\partial_xUU^\dagger \Lambda_2U\partial_xU^\dagger \Lambda_1
  +\partial_xUU^\dagger \Lambda_3U\partial_xU^\dagger \Lambda_2+\partial_xUU^\dagger \Lambda_1U\partial_xU^\dagger \Lambda_3\right] \\
&+
{1\over 4aJ} \hbox{tr}\left[\partial_\tau UU^\dagger \Lambda_2U\partial_\tau U^\dagger \Lambda_1
  +\partial_\tau UU^\dagger \Lambda_3U\partial_\tau U^\dagger \Lambda_2+\partial_\tau UU^\dagger \Lambda_1U\partial_\tau U^\dagger \Lambda_3\right],
\end{split}
\end{equation}
where the $\Lambda_i$ matrices are defined in Eq.~\eqref{defLambda}.  To get the perturbative spectrum
we can expand $U$ in the Gell-Mann matrices:
\be U=e^{i\vec \theta \cdot \vec T}\approx I+i\vec \theta \cdot \vec T.\ee
Only the off-dagonal Gell-Mann matrices appear in the Lagrangian. 
Then to quadratic order:
\begin{equation}
\begin{split}
{\cal L}_{\hbox{real}} \approx \ &{}ap^2J \left[4(\partial_x\theta_1)^2+4(\partial_x\theta_2)^2
  +(\partial_x\theta_4)^2+(\partial_x\theta_5)^2+(\partial_x\theta_6)^2+(\partial_x\theta_7)^2\right] \\
&+{1\over 4aJ}\left[(\partial_\tau \theta_1)^2+(\partial_\tau \theta_2)^2+(\partial_\tau \theta_4)^2
  +(\partial_\tau \theta_5)^2+(\partial_\tau \theta_6)^2+(\partial_\tau \theta_7)^2\right].
\end{split}
\end{equation}
We see that the velocities of the 6 perturbative Goldstone modes are:
\be v_1=v_2=4apJ,\ \  v_4=v_5=v_6=v_7=2apJ\ee
consistent with the flavour-wave theory results of Eq.~\eqref{eq:su3-hw-v}.
The different velocities implies that the model is not Lorentz invariant. Due to the Mermin-Wagner-Coleman
theorem,\cite{MerminWagner1966,Coleman1973} we expect that the $\mbox{SU}(3)$ symmetry will not be spontaneously broken once interaction effects
are taken into account and no Goldstone modes will appear in the actual spectrum. Note that only low energy excitations appear in the path integral, therefore the flat modes don't contribute due to the exponential cutoff in energy. 

\subsection{Symmetries of the NLI$\sigma$M}
\label{sec:symmetries}

As we can see in Eq.~\eqref{eq:symmaction}, we arrive at an $\mbox{SU}(3)/( \mbox{U}(1) \times \mbox{U}(1))$ non-Lorentz invariant flag manifold $\sigma$-model (NLI$\sigma$M), similarly to the case of $\mbox{SU}(3)$ chains in the fully symmetric irrep.\cite{LajkoNuclPhys2017} However, the origin of the fields is different in the two cases. In the fully symmetric case, the classical ground state had a three-sublattice order and the  $\vec \phi^1, \vec \phi^2, \vec \phi^3$  correspond to the three spin states inside a unit cell. 
On the other hand, in the self-conjugate case, the classical ground state is a two-sublattice ordered state, and only $\vec \phi^1,\vec \phi^2$ correspond to spin states directly, while $\vec \phi^3$ is uniquely defined from the other two. As a result the symmetries of the underlying spin models give rise to different symmetries in the field theory. Here we  go through these symmetries in the self-conjugate case; the symmetries of the fully symmetric case can be found in Sec.~5  of Ref.~\onlinecite{LajkoNuclPhys2017}.  

Assuming $\mbox{SU}(3)$, gauge and time reversal invariance the general form of the $\mbox{SU}(3)/( \mbox{U}(1) \times \mbox{U}(1))$ NLI$\sigma$M is  \cite{LajkoNuclPhys2017}

\bea
S&=&  \int dx d\tau  \Bigg(   \Big[\frac{v_{1,2}}{g_{1,2}}|\vec \phi^{2*}\cdot\partial_x \vec \phi^1|^2+  \frac{v_{2,3}}{g_{2,3}} |\vec \phi^{3*}\cdot\partial_x \vec \phi^2|^2 
+  \frac{v_{3,1}}{g_{3,1}}|\vec \phi^{1*}*\cdot\partial_x \vec \phi^3|^2\Big]
\nonumber \\
&+& \Big[ \frac{1}{v_{1,2}g_{1,2}}| \vec \phi^{2*}\cdot\partial_\tau \vec \phi^1|^2+ \frac{1}{v_{2,3}g_{2,3}}| \vec \phi^{3*}\cdot\partial_\tau \vec \phi^2|^2
+\frac{1}{v_{3,1}g_{3,1}}| \vec \phi^{1*}\cdot\partial_\tau \vec \phi^3|^2\Big]\Bigg)\nonumber \\
&+&i\big( \theta_1 Q_1 +\theta_2 Q_2\big) +i\lambda (q_{12}+q_{23}+q_{31}),
\label{eq:genaction}
 \eea
 where the imaginary $\lambda$-term is discussed in detail in Refs.\ \onlinecite{LajkoNuclPhys2017, OhmoriSeiberg2019}.
In the specific case of the nearest neighbour self-conjugate Heisenberg Hamiltonian, the action obtained in Eq.~\eqref{eq:symmaction} corresponds to $v_{1,2} =4apJ$, $v_{2,3}=v_{3,1}= 2apJ$, $g_{1,2}=1/p$, $g_{2,3}=g_{3,1}=2/p$,  $\theta_1=-\theta_2=\pi$ and $\lambda=0$. 
In the following we go through the symmetries of the Hamiltonian in Eq.~\eqref{eq:su3-hw-H} and examine what symmetries they give on the parameters of the NLI$\sigma$M.
\subsubsection{Translation by one site}

Under translation by one site the $\vec \phi^1(x,\tau)$ and $\vec \phi^2(x, \tau)$ fields map to each other, while $\vec \phi^3 := \vec \phi^{1*}\times \vec \phi^{2*}$ maps to $-\vec \phi^3 = \vec \phi^{2*}\times \vec \phi^{1*}$.  This transformation maps the $g_{2,3}$ and $g_{3,1}$ terms  to each other, therefore requiring that $v_{2,3}=v_{3,1}$ and $g_{2,3}=g_{3,1}$. 

The $\lambda$-term transforms as

\be
\lambda (q_{12}+q_{23}+q_{31}) \to \lambda (q_{21}+q_{13}+q_{32}) = -\lambda (q_{12}+q_{23}+q_{31}) \ee
which is only invariant if $\lambda =0$ (note that the $q_{12}+q_{23} +q_{31}$ term is not integer valued, thus $\lambda =0$, not only mod $2\pi$). 
As for the topological term,  $\theta_1 Q_1+\theta_2 Q_2$ maps to $\theta_1 Q_2 +\theta_2 Q_1$, thus guaranteeing $\theta_1=\theta_2$ mod $2\pi$. It is interesting to note that this doesn't fix $\theta_1 ,\theta_2  = \pi \text{ or }  0$ on its own. 

We find that the translational invariance of the  self-conjugate $\mbox{SU}(3)$  chain maps to a $\mathbb{Z}_2^{(\text{tr})}$ symmetry  (`tr' stands for translation) of the NLI$\sigma$M; a translation by two lattice sites would map to the identity transformation of the NLI$\sigma$M. This is the consequence of the two-sublattice ordered classical ground state.  In the case of the fully symmetric $\mbox{SU}(3)$ chain, the translational invariance results in a $\mathbb{Z}_3$ symmetry, which corresponds to the cyclic permutation of the three fields.  This is because the classical ground state has a three-sublattice structure in that case. The absence of this $\mathbb{Z}_3$ symmetry in the current case is manifested in the different coupling constants in the action, and has  important consequences on the phase diagram, as we will discuss later.

\subsubsection{Site Parity}

The parity symmetry around a site maps each sublattice to itself, and inverts the position of the spins, therefore at the level of the NLI$\sigma$M it maps to a $\mathbb{Z}_2^{(sp)}$ symmetry taking $\vec \phi^i (x,\tau)$ to $\vec \phi^i (-x,\tau)$ (`sp' stands for site parity).  Under this symmetry the real part of the action remains invariant, so it doesn't give any constraint on the velocities or coupling constants. 
The $\lambda$-term and the topological term both get a minus sign, since they always contain exactly  one spatial derivative. Therefore this symmetry is only satisfied if  $\lambda=0$, and if $\theta_1, \theta_2= 0$ or $\pi \text{ mod } 2\pi$, but in itself doesn't fix the two angles to be equal. 

\subsubsection{Bond Parity}

The parity symmetry around a bond centre maps the two sublattices into each other and flips the spatial coordinate. At  the level of the NLI$\sigma$M this corresponds to another $\mathbb{Z}_2^{(\text{bp})}$ symmetry (`bp' stands for bond parity)  mapping $\vec \phi^1 (x,\tau) \to  \vec \phi^2(-x,\tau)$,
$\vec \phi^2 (x,\tau) \to  \vec \phi^1(-x,\tau)$ and $ \vec \phi^3(x,\tau) \to -\vec \phi^3(-x,t)$.  This symmetry once again fixes  $v_{2,3}=v_{3,1}$ and $g_{2,3}=g_{3,1}$. 
The $\lambda$-term transforms as  ${q_{12} +q_{23}+q_{31}} \to{  -q_{21} - q_{13}-q_{32}}$, therefore  it is invariant for any $\lambda$ due to Eq.~\eqref{eq:qmn}. The topological term transforms as $ \theta_1 Q_1 + \theta_2 Q_2\to -\theta_1 Q_2 - \theta_2 Q_1$, which is invariant  for any $\theta_1 = -\theta_2$.

\subsubsection{$a_\alpha \leftrightarrow b^\alpha$ invariance and charge conjugation}

In the self-conjugate  $\mbox{SU}(3)$  chain model interchanging the role of the two types of bosons
\be a_{\alpha ,n}\leftrightarrow b^\alpha_n.\ee
 transforms the spin operators as
\be S^\alpha_\beta =a^{\alpha \dagger} a_\beta -b^\dagger_\beta b^\alpha \to
b_\alpha^\dagger b^\beta -a^{\beta \dagger}a_\alpha = -S^\beta_\alpha ,\ee
which is clearly a symmetry of the Hamiltonian. In the coherent state language this corresponds to
\be z_{\alpha ,n}\leftrightarrow w^\alpha_n\ee
or, equivalently
\be \Phi^1_\alpha\leftrightarrow \Phi^{2\alpha *},\ee
which in the field theory translates to the $\mathbb{Z}_2^{(a\leftrightarrow b)}$ symmetry 
\be \vec \phi^1 (x,t)\leftrightarrow \vec \phi^{2 *}(x,t),  \quad \vec \phi^3 (x,t)\rightarrow - \vec \phi^{3 *}(x,t).
\ee
This symmetry has similar consequences as the translation invariance, namely it guarantees that $v_{2,3}=v_{3,1}$, $g_{2,3}=g_{3,1}$ and $\lambda=0$,  but forces $\theta_1 =-\theta_2$, since $Q_1$ is mapped to $-Q_2$ ( because of the complex conjugation) and vice versa.

If we combine this symmetry with translation by one site, we get charge conjugation in the field theory
\be \vec \phi^i (x,t)\leftrightarrow \vec \phi^{i*} (x,t).\ee

\subsection{Breaking the symmetries}
\label{sec:symmbreak}

Here we briefly discuss how one can break the above symmetries in the self-conjugate  $\mbox{SU}(3)$  chain model by introducing dimerized nearest or next nearest neighbour couplings. Note, however, that the $ a_{\alpha ,n}\leftrightarrow b^\alpha_n$ symmetry cannot be broken unless we break the fundamental  $\mbox{SU}(3)$  symmetry, or  consider other, non self-conjugate representations. As a result, the currently considered spin models will always map to NLI$\sigma$Ms with $\theta_1=-\theta_2$, $\lambda=0$, and  $v_{2,3}=v_{3,1}$, $g_{2,3}=g_{3,1}$. 
Actually, the ratio between the velocities $v_{1,2}$ and $v_{2,3}=v_{3,1}$ is also fixed at 2, independently of the breakdown of the lattice symmetries. This is a consequence of the self-conjugate irreps and $\mbox{SU}(3)$  symmetry, as discussed in more detail in the flavour-wave approach in Appendix~\ref{app:modes}.  %Therefore the resulting family of NRSMs is much smaller than for $\mbox{SU}(3)$  chains in the fully symmetric irrep, 

Considering a self-conjugate $\mbox{SU}(3)$ model with alternating $J_1$ and $J_1'$ nearest neighbour, and alternating $J_2$  and $J_2'$  next nearest neighbour interactions, the resulting $\sigma$-model reads as:
\begin{equation}
\begin{split}
S=  \int dx d\tau  \Bigg( & \frac{v}{g} \Big[4|\vec \phi^{2*}\cdot\partial_x \vec \phi^1|^2+  |\vec \phi^{3*}\cdot\partial_x \vec \phi^2|^2 +|\vec \phi^{1*}\cdot\partial_x \vec \phi^3|^2\Big]
 \\
&+ \frac{1}{vg}\Big[ | \vec \phi^{2*}\cdot\partial_\tau \vec \phi^1|^2+| \vec \phi^{3*}\cdot\partial_\tau \vec \phi^2|^2+ | \vec \phi^{1*}\cdot\partial_\tau \vec \phi^3|^2\Big]  \Bigg) \\
&+i\theta \big( Q_1 - Q_2\big), \end{split}
\label{eq:dimaction}
\end{equation}
where
\begin{equation}
\begin{split}
\frac{v}{g}= 2ap^2\left(\frac{J_1J_1'}{J_1+J_1'} - (J_2+J_2')\right), \quad \frac{1}{vg} = \frac{1}{2a (J_1+J_1')}, \quad  \theta=\frac{2p\pi J_1'}{J_1+J_1'}
\end{split}
\label{eq:dimactioncoeffs}
\end{equation}
which gives a coupling constant $1/g = p\sqrt{J_1J_1' -(J_2+J_2')(J_1+J_1')}/(J_1+J_1')$ (the velocity $v$ can be set to $1$ by rescaling the space and time variables). Any longer range coupling would be equivalent to the nearest or next nearest couplings at the level of the NLI$\sigma$M.  %Also, if we considered models which doesn't have two-site translational invariance, the whole path integral derivation would change and we would get a different family of NLSMs. 
 If $J_1=J_1'$, the site parity is conserved independently of $J_2,J_2'$, thus fixing the topological angle to $p\pi$ (together with the  $a_\alpha \leftrightarrow b^\alpha$ invariance). The difference between $J_2, J_2'$ has no effect on the underlying theory and the next nearest neighbour interactions only rescale the coupling constants and velocities, allowing us to tune the coupling constant and drive the system to $g \to \infty$, without changing the topological term. The two-sublattice classical ground state is only stable for $J_2<J_1/4$. For larger $J_2$ the classical ground state becomes helical, and thus our two-sublattice path integral approach breaks down, which is manifested in the diverging coupling constant $g$. 
 
Breaking the translational invariance and site parity by introducing dimerization in the nearest neighbour bonds tunes the topological angle away from $p\pi$, while keeping $\theta_1=-\theta_2$. For $J_1'=J_1/2$ we reach the   $\theta_1=-\theta_2=2p\pi/3$ point, where the theory is invariant under the above mentioned $\mathbb{Z}_3$ transformation.  As  was discussed in Ref.\ \onlinecite{LajkoNuclPhys2017}, in the infinite coupling limit this point corresponds to a first order phase transition point with spontaneously broken $\mathbb{Z}_3$ symmetry. For NLI$\sigma$Ms where all couplings are the same the action possesses this $\mathbb{Z}_3$ even for finite couplings. In those models, for strong but finite coupling the $\mathbb{Z}_3$ remains spontaneously broken only until a critical coupling $g_c$ below which the $\theta_1=-\theta_2=2\pi/3$ point becomes a gapless critical point. However, in case of the self-conjugate irreps the couplings are not equal, $g_{1,2}=2g_{1,3}=2g_{2,3}$ and the $\mathbb{Z}_3$ is explicitly broken for any finite coupling. %In Sec.~\ref{sec:MC} we will show that this causes the disappearance of the critical point present in the fully symmetric case.

\section{\label{sec:LSMA}Failure of LSMA theorem}
For the symmetric irreps of  $\mbox{SU}(3)$  the LSMA theorem can be proven \cite{AffleckLieb1986,  LajkoNuclPhys2017} by acting on a ground state, for a chain of length $L$ with periodic boundary conditions, with the unitary operator
\be U := \exp [i\sum_j (2\pi j/3L)\fQ_j]\ee
with 
\be \fQ_j := S^1_{1,j}+S^2_{2,j}-2S^3_{3,j}.\ee
Then under translation by one site
\be TUT^\dagger =Ue^{-i(2\pi /3L)\fQ}e^{i(2\pi /3)\fQ_1},\ \  \fQ := \sum_j\fQ_j.\ee
Since the ground states obey $\fQ|\psi_0\rangle =0$, we have
\be TU|\psi_0\rangle =e^{i(2\pi /3)\fQ_1}U|\psi_0\rangle \ee
where the ground state $|\psi_0\rangle$ was chosen to obey $T|\psi_0\rangle =|\psi_0\rangle$. 
For the fully symmetric irrep with $p$ boxes, 
\be e^{i(2\pi /3)\fQ_1}|\psi_0\rangle=e^{i(2\pi /3)p}|\psi_0\rangle .\ee
Thus $TU|\psi_0\rangle =e^{i(2\pi /3)p}U|\psi_0\rangle$, 
\be \langle \psi_0|[U^\dagger HU-U]|\psi_0\rangle =\mathcal{O}(1/L),\ee
and by considering translation by one site
\be \langle \psi_0|U|\psi_0\rangle =e^{i(2\pi /3)p}\langle \psi_0|U|\psi_0\rangle =0\ \  (p\neq 3m).\ee
Thus $U|\psi_0\rangle$ is a low energy state orthogonal to $|\psi_0\rangle$, for $p\neq 3m$, implying either gapless excitations or spontaneously broken translation symmetry. 
 On the other hand, for the self-conjugate representations, 
\be \fQ=a^{1\dagger}a_1+a^{2\dagger}a_2-2a^{3\dagger}a_3-b^\dagger_1b^1-b^\dagger_2b^2+2b^\dagger_3b^3
=3(b^\dagger_3b^3-a^{3\dagger}a_3)\ee
where we used $(\vec a^\dagger \cdot \vec a-\vec b^\dagger \cdot \vec b)|\psi_0\rangle =0$. So
\be e^{i(2\pi /3)\fQ}=e^{i2\pi (b^\dagger_3b^3-a^{3\dagger}a_3)}=1,\ee
where we used the fact that $a^{3\dagger}a_3$ and $b^\dagger_3b^3$ have integer eigenvalues. So, translating 
by one site maps $U|\psi \rangle$ into $U|\psi \rangle$ with no phase for any value of $p$. Thus 
there is no proof that $U|\psi \rangle$ is orthogonal to $|\psi \rangle$ so the LSMA theorem fails.
Actual models with  short range interactions and  conserved  $\mbox{SU}(3)$  symmetry and a unique gapped ground state exist for odd $p$,\cite{Gozel2019} giving direct evidence against a possible LSMA theorem. 

%%%%%%%%%%%%%
%%%%%%%%%%%%%
%%%%%%%%%%%%%

\section{Strong coupling limit of field theory}
\label{sec:strongcoupling}
In the strong coupling limit, the real terms in the action vanish and only
the topological terms remain. In this limit, Lorentz invariance is restored.
The action is simply
\be S=i\pi (Q_1-Q_2)\ee
for $p$ odd. Following the techniques of Refs.\ \onlinecite{SeibergPRL1984,PlefkaSamuelPRD1997} this limit was
solved using a lattice formulation in Ref.\ \onlinecite{LajkoNuclPhys2017}. The partition function,
for arbitrary topological angles, becomes:
\be Z(\theta_1,\theta_2 )\to \sum_{m,n\in Z}z(\theta_1+2\pi m, \theta_2+2\pi n)^A,\ee
where
$A$ is the area of the 2-dimensional space-time (divided by the area of a plaquette
in the lattice model) and
\be z(\theta_1,\theta_2)=2{(\theta_1-\theta_2)\cos \left({\theta_1-\theta_2\over 2}\right)
  -\theta_1\cos \left({\theta_1\over 2}\right)
  + \theta_2\cos \left({\theta_2\over 2}\right)
\over \theta_1\theta_2(\theta_1-\theta_2)}.\ee
In the infinite area limit the sum is dominated by the values of $m$ and $n$ which give
the largest value of $|z|$.  For $\theta_1$ near $-\pi$ and $\theta_2$ near $\pi$, it can
be seen that
\be \max_{m,n}z(\theta_1+2\pi m, \theta_2+2\pi n)\approx {2\over \pi^2}+{1\over \pi}
+{2|\theta_1+\theta_2|\over \pi^3}\ee
with the dominant $(m,n)$ terms being
$(m,n)=(1,0)$ for $\theta_1+\theta_2<0$ and $(m,n)=(0,-1)$ for
 $\theta_1+\theta_2>0$.
The expectation value of the topological charges can be written
\be\langle Q_i\rangle =iA{\partial \ln z\over \partial \theta_i}.\ee
Approaching the line $\theta_1=-\theta_2$ from the side $\theta_1+\theta_2>0$ or
$\theta_1+\theta_2<0$ we get 2 different results:
\be\langle Q_1\rangle=\langle Q_2\rangle =\pm iA{2\over \pi (2+\pi )}.
\label{eq:symmbreakQ}
\ee
This is indicative of a first order phase transition along the line $\theta_1=-\theta_2$, which corresponds to the breakdown of the bond parity ($n\to 1-n$) and the $a_\alpha\leftrightarrow b^\alpha$ parity of the spin model.  Both of these take  $\theta_1 \leftrightarrow -\theta_2$ in the field theory,  therefore they are exact symmetries for $\theta_1=-\theta_2 =\pi$. But they also map $ \langle Q_1\rangle \leftrightarrow -\langle Q_2 \rangle$, therefore  the topological charge averages in Eq.~\eqref{eq:symmbreakQ} clearly show that these symmetries are spontaneously broken in the strong coupling limit. 

It is plausible that broken parity symmetry occurs for $\theta_2=-\theta_1=\pi$ even for weak coupling.  Indeed that is consistent with the renormalization group (RG) flow diagram suggested in Fig.~1b) of Ref.\ 
\onlinecite{LajkoNuclPhys2017}. While critical points are expected at finite coupling at $\theta_1=-\theta_2=\pm 2\pi /3$, corresponding to the $\mbox{SU}(3)_1$ Wess-Zumino-Witten model \cite{LajkoNuclPhys2017, Sulejmanpasic2018,OhmoriSeiberg2019} no such 
critical points are expected at $\theta_1=-\theta_2=\pm \pi$. In that case we may expect an RG flow from weak coupling to strong coupling where broken parity occurs. A further complication 
is the breaking of Lorentz invariance and of the $\mathbb{Z}_3$ symmetry cyclically exchanging the three $\vec \phi^i$ fields.  Both these symmetries are present in the field theory studied in Ref.\ \onlinecite{LajkoNuclPhys2017}, arising from chains in the fully symmetric irrep, but not in the field theory studied here.

\section{Monte Carlo simulations}
\label{sec:MC}

Refs.\ \onlinecite{Sulejmanpasic2018, OhmoriSeiberg2019} use a  't Hooft anomaly argument \cite{tHooft1980} to predict a gapless or trimerized behaviour at  $\theta_1=-\theta_2=2\pi/3$ in the {\it Lorentz invariant} sigma model with equal couplings.  This anomaly argument relies on the presence of a  $\mathbb{Z}_3$ symmetry corresponding to the cyclic permutation of the fields. In the current case this symmetry  is explicitly broken at finite coupling for any values of the  topological angles due to the different coupling constants.  As a result  there is no anomaly at $\theta_1=-\theta_2=2\pi/3$, or at any other value of topological angles, in agreement with the failure of the LSMA theorem. 

We carried out Monte Carlo simulations to study the fate of the $\mbox{SU}(3)_1$ critical point at  $\theta_1=-\theta_2=2\pi/3$ when one of the coupling constants is tuned away from the isotropic case.   As it is discussed in Appendix E of Ref.~\onlinecite{LajkoNuclPhys2017}, the action of  Eq.~\eqref{eq:symmaction} can still be rewritten as three copies of a $\mbox{CP}^2$ theory, even for unequal coupling constants and velocities. Therefore the real part on the lattice can be written as 
\begin{equation}
\begin{split}
S^{(\text{lattice})}_{\text{real}}= -\sum_{\vec r} \Bigg[&\mathbin{\phantom{+}}\frac{v}{2g}\bigg(  \left| \vec\phi^{1*} (\vec r)\cdot  \vec\phi^1( \vec r + \vec \delta_x)\right|^2+\left| \vec\phi^{2*} (\vec r)\cdot  \vec\phi^2( \vec r + \vec \delta_x)\right|^2+\alpha \left| \vec\phi^{3*} (\vec r)\cdot  \vec\phi^3( \vec r + \vec \delta_x)\right|^2 \bigg) \\
&+ \frac{1}{2vg}\bigg(  \left| \vec\phi^{1*} (\vec r)\cdot  \vec\phi^1( \vec r + \vec \delta_\tau)\right|^2+\left| \vec\phi^{2*} (\vec r)\cdot  \vec\phi^2( \vec r + \vec \delta_\tau)\right|^2+\left| \vec\phi^{3*} (\vec r)\cdot  \vec\phi^3( \vec r + \vec \delta_x)\right|^2\bigg) \Bigg]
\end{split}
\label{eq:discretizedaction_diag}
\end{equation}
where  $v=4apJ$, $g=1/p$, and  $\alpha=-1/2$. The difference in the coupling constants and velocities in  Eq.~\eqref{eq:symmaction}  manifests in the $\alpha$ parameter. The topological term on the lattice is written following the recipe of Berg and L\"uscher,\cite{BergLuscher1981} which guarantees that the topological charges are integer valued even in the discretized system action. 

  We can make Monte Carlo simulations for this lattice action for imaginary angles when the topological term is real.\cite{Azcoiti2012,AllesPapa2008,AllesPapa2014} We set $v=1$, and changed the values of $g$ for $\alpha=-1/2$. We used a multigrid update method \cite{Hasenbuschmultigrid, LajkoNuclPhys2017} to decrease autocorrelations.   For each imaginary angle and $\alpha$ we sampled $5 \times10^4$  configurations with a sampling distance of 10 multigrid sweeps after $5 \times10^4$ thermalizing multigrid sweeps.  We  obtained the mass gap from the inverse of the correlation length.  We then extrapolated the mass gap values from imaginary to real angles by fitting a function of the form $(c_1 + c_2\theta^2)/(1 + c_3\theta^2)$.   As it was discussed in Ref.\ \onlinecite{LajkoNuclPhys2017}, we fitted values until the inflection point in the mass gap results, beyond which  there is a change in behaviour due to saturation of the topological charge density.\cite{ImachiYoneyamaPTP2006} 

 \begin{figure}[tb]
\begin{center}
\includegraphics[width=0.49\textwidth]{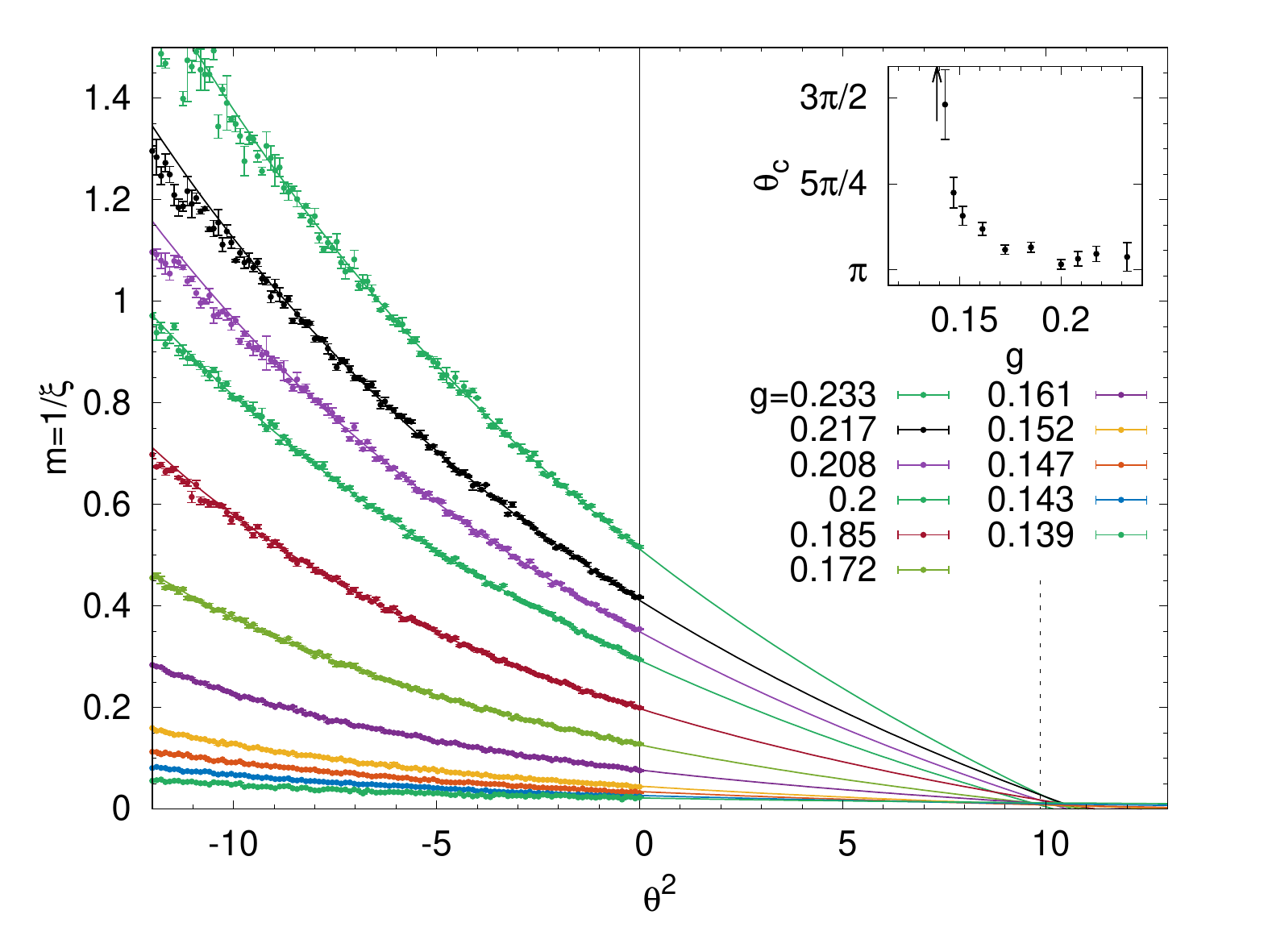}
\caption{Extrapolating the inverse of the correlation length along the $\theta_1=-\theta_2$ line, based on simulations on a $192 \times 192$ system for various couplings and fixed $\alpha=-1/2$,  which corresponds to the self-conjugate irreps. The results suggest that the system remains gapped even at $\theta=\pi$, and there is no gapless transition for any $\theta<\pi$. }
\label{fig:alphatune}
\end{center}
\end{figure}

% In Fig.~\ref{fig:alphatune}a we present results for a fixed coupling constant, and different  values of $\alpha$. %from $1$ to $-1/2$, which interpolates between the fully symmetric and the self-conjugate cases. 
 %We selected $g=0.233$, so that for the  isotropic $\alpha=1$ case the $\theta_1=-\theta_2=2\pi/3$ point is still in the gapless phase  ($g_c \approx 0.255$)\cite{LajkoNuclPhys2017}.  We find that as soon as we tune away from $\alpha=1$ the gap at $2\pi/3$ opens up, or from  another point of view the extrapolated gap closes beyond $2\pi/3$. Interestingly this seems to be the case for both $\alpha<1$ and $\alpha >1$.  
The results  are shown in Fig. \ref{fig:alphatune}. We find that the extrapolated gap always closes beyond $\theta=\pi$; since $\theta$ and $-\theta$ are equivalent, this suggests that the gap stays finite for all $\theta$ and for all values of g. Note that this extrapolation can detect a gapless phase transition thanks  to the closing of the extrapolated mass gap, but it cannot detect if a gapped transition happens. If the latter is the case, the extrapolated values beyond the transition are not physical, so even if the extrapolated gap closes further on, it doesn't mean that there is an actual gapless point. What we can tell  is that there can't be any gapless transition before the extrapolated gap closes. Based on the strong coupling considerations we believe that there is a first order phase transition for some $\theta_c$ between $0$ and $\pi$ separating a trivial phase for $\theta<\theta_c$ from a gapped doubly degenerate phase with spontaneous parity breaking for $\theta_c<\theta$.  In Appendix \ref{app:symmbreak} we discuss  the possibility to detect  spontaneous symmetry breaking by measuring various order parameters.  Unfortunately so far we didn't manage to extract meaningful results, due to difficulties in the extrapolation, therefore the verification of the proposed transition requires further study.

\section{Exact ground states}
\label{sec:AKLT}

\begin{figure}[t]
\begin{center}
\includegraphics[width=0.9\textwidth]{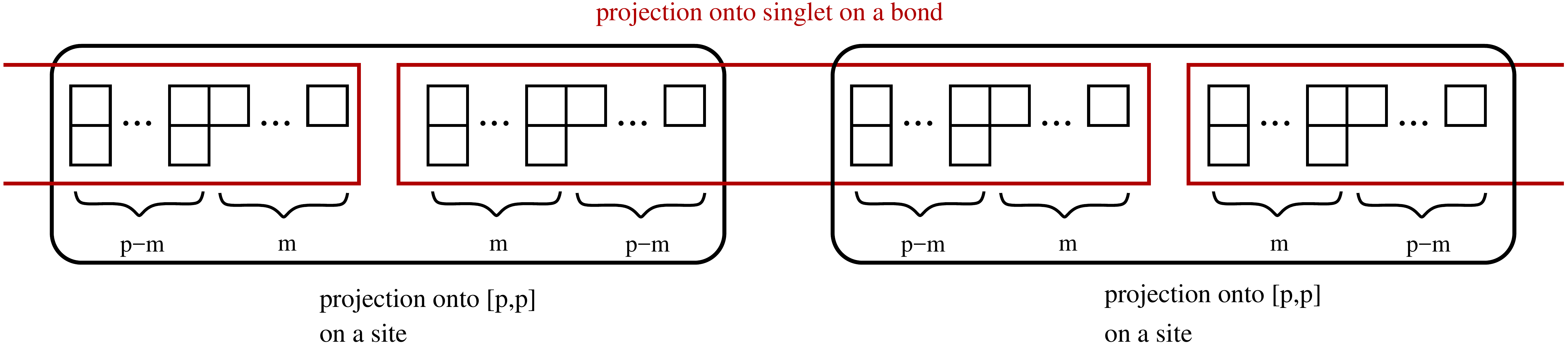}
\caption{Family of  AKLT states  for the $[p,p]$ self-conjugate spin state. Each spin is split into a $[m,p-m]$ and a $[p-m,m]$ virtual spin. Then each of these virtual spins form singlets with the other type of virtual spin on the neighbouring site. On each site the tensor product of the virtual spins is projected onto the $[p,p]$ irrep. For $p$ odd, the states with $m=(p-1)/2$ and $m=(p+1)/2$ correspond to the states in Eq.~\ref{eq:oddAKLT}. They are connected to each other by the bond parity, therefore forming doubly degenerate ground states of the appropriate parent Hamiltonian. For $p$ even, the $m=p/2$ state is invariant under bond parity, giving a unique ground state. }
\label{fig:AKLT}
\end{center}
\end{figure}

Finally we can write down simple states, analogous to the AKLT states.\cite{AKLT1988}  Let's write the states of the $[p,p]$ irrep on a single site as 
\be
 \ket{\begin{array}{l} {\alpha_1\alpha_2\alpha_3\ldots \alpha_p}\\{\alpha_{p+1}\alpha_{p+2}\alpha_{p+3}\ldots \alpha_{2p}}\end{array}} = \mathcal{P}_{[p,p]} \big( a^{\alpha_1\dagger}a^{\alpha_2\dagger}\dots a^{\alpha_p\dagger} b_{\alpha_{p+1}}^\dagger b_{\alpha_{p+2}}^\dagger \dotsb_{\alpha_{2p}}^\dagger \ket{0}\big),
\ee
where $\mathcal{P}_{[p,p]}$ projects the state with p $a$ bosons and p $b$ bosons onto the $[p,p]$ irrep. 
These are symmetric under interchanging raised or lowered indices and traceless under contracting any pairs of upper and lower indices.\cite{MathurSen2001} Then for $p$ even the AKLT state on the bond between sites $n$ and $n+1$ reads as
\be
\dots \otimes \ket{ 
\begin{array}{r} {\alpha_1\ldots \alpha_{p/2}}\\{\alpha_{p/2+1}\ldots \alpha_{p}}\end{array} \begin{array}{l} \beta_1\ldots \beta_{p/2} \\{\beta_{p/2+1}\ldots \beta_{p}}\end{array} }_n \otimes \ket{ \begin{array}{r}{\beta_{p/2+1}\ldots \beta_{p}}\\ {\beta_1\ldots \beta_{p/2}}\end{array}    \begin{array}{l} {\gamma_{1}\ldots \gamma_{p/2}}\\
{\gamma_{p/2+1}\ldots \gamma_{p}} \end{array}}_{n+1}\otimes\dots
\ee
Note that $p/2$ upper indices on site $n$ are contracted with $p/2$ lower indices on site $n+1$ and $p/2$ lower indices on site $n$ are contracted with $p/2$ upper indices on site $n+1$.
For $p$ odd there are two AKLT states of the form

\begin{equation}
\begin{split}
\dots \otimes \ket{ 
\begin{array}{r} {\alpha_1\ldots \alpha_{(p-1)/2}}\\{\alpha_{(p+1)/2}\ldots \alpha_{p}}\end{array} \begin{array}{l} \beta_1\ldots \beta_{(p+1)/2} \\{\beta_{(p+3)/2}\ldots \beta_{p}}\end{array} }_n \otimes \ket{ \begin{array}{r}{\beta_{(p+3)/2}\ldots \beta_{p}}\\ {\beta_1\ldots \beta_{(p+1)/2}}\end{array}    \begin{array}{l} {\gamma_{1}\ldots \gamma_{(p+1)/2}}\\
{\gamma_{(p+3)/2}\ldots \gamma_{p}} \end{array}}_{n+1}\otimes\dots\\
\dots \otimes \ket{ 
\begin{array}{r} {\alpha_1\ldots \alpha_{(p+1)/2}}\\{\alpha_{(p+3)/2}\ldots \alpha_{p}}\end{array} \begin{array}{l} \beta_1\ldots \beta_{(p-1)/2} \\{\beta_{(p+1)/2}\ldots \beta_{p}}\end{array} }_n \otimes \ket{ \begin{array}{r}{\beta_{(p+1)/2}\ldots \beta_{p}}\\ {\beta_1\ldots \beta_{(p-1)/2}}\end{array}    \begin{array}{l} {\gamma_{1}\ldots \gamma_{(p-1)/2}}\\
{\gamma_{(p+1)/2}\ldots \gamma_{p}} \end{array}}_{n+1}\otimes\dots
\end{split}
\label{eq:oddAKLT}
\end{equation}
%
%
%
%\be |{}^{\alpha_1\ldots \alpha_{(p-1)/2}\alpha_{(p+1)/2+1}\ldots \alpha_p}_{\alpha_{p+1}\ldots \alpha_{(3p-1)/2}\alpha_{((3p+1)/2+1}\ldots \alpha_{2p}},{}^{\alpha_{(3p-1)/2)}\ldots \alpha_{2p}\alpha_{2p+1}\ldots \alpha_{(5p-3)2}}
%_{\alpha_{(p+1)/2}\ldots \alpha_{p}\alpha_{(5p-1)/2}\ldots \alpha_{3p}},\ldots >\ee
In one of the AKLT states, $(p+1)/2$ upper indices and $(p-1)/2$ lower indices on site $n$ are contracted with site $n+1$. In the other AKLT state 
$(p-1)/2$ upper indices and $(p+1)/2$ lower indices on site $n$ are contracted with site $n+1$. Both of these states are translation invariant and map into each other under the bond parity or the $a \leftrightarrow b$ parity as well. 
Fig.~\ref{fig:AKLT} provides an illustration of this construction. For $p=1$, this is similar to the construction of  Morimoto et al.\cite{MorimotoFurusakiPRB2014}

With this type of construction we can't find any AKLT type states, for $p$ odd, which do not break bond parity and the $a_\alpha \leftrightarrow b^\alpha$ parity,  which is consistent with our conjecture that it is spontaneously broken. For  $p$ odd and any given $0\leq m\leq p$ the state depicted in Fig.~\ref{fig:AKLT} is connected to the state with $m'=p-m$ by bond parity or $a_\alpha \leftrightarrow b^\alpha$ parity.  This is even true for  $p$ even for any $m\neq p/2$. 
As was discussed recently in Ref.~\onlinecite{Gozel2019}, Hamiltonians can be found for which these are the unique (or doubly degenerate) exact ground states. 

\section{Conclusions}
\label{sec:conclusions} 
We have studied the self-conjugate representations of SU(3) whose Young Tableaux contain p columns of
length 2 and p columns of length 1, focusing on the large-$p$ limit. We mapped the ``spin'' chains into a
non-Lorentz invariant $\sigma$-model, for large $p$.  We obtained topological angles in the field theory
which are 0 for $p$ even and $\pi$ for $p$ odd.
The field theory is not Lorentz invariant due to unequal velocities for different
perturbative excitations.
We have confirmed the perturbative limit of the field theory using
flavour-wave theory. 

Based on our proposed phase diagram for the Lorentz invariant version of this field theory
in Ref.~\onlinecite{LajkoNuclPhys2017},
our analysis of the strong coupling limit of the field theory
and AKLT states that we constructed, 
we conjecture a gapped phase for all $p$, with spontaneously broken parity and $a_\alpha \leftrightarrow b^\alpha$ symmetries for $p$-odd only.
The gap was confirmed using a Monte Carlo study of the field theory. However, we were unable to
confirm the broken symmetries using Monte Carlo simulations.  AKLT states can be constructed \cite{Gozel2019} 
which do not have broken parity symmetry. So, for sufficiently general Hamiltonians, ground states
without broken parity exist for $p$ odd. The Lieb-Schultz-Mattis-Affleck theorem, which
proves either broken symmetry or a gapless ground states for the p-box  symmetric representations with $p\neq 3$,
fails for the self-conjugate representations. Nonetheless, we conjecture broken symmetry when $p$ is odd for
the simplest Hamiltonian, which we consider here. Our broken symmetry conjecture definitely
needs further confirmation. In  Appendix \ref{app:symmbreak} we present possible order parameters that could be used to detect these symmetry breakings, and briefly discuss our attempts to demonstrate it, in the hope that it might be useful to the community.

\section*{Acknowledgements} 
We would like to thank Nathan Seiberg for helpful communications. IA is supported by NSERC of Canada Discovery Grant
04033-2016 and by the Canadian Institute for Advanced Research. KW is supported by an NSERC PGS-D and the QuEST program
of the Stewart Blusson Quantum Matter Institute. FK, ML and FM gratefully acknowledge the financial support of the Swiss
National Science Foundation. The numerical simulations have been performed using the facilities of the Scientific IT and Application Support Center of EPFL.

\appendix
\section{Traceless condition in the Mathur-Sen representation}
\label{appendixA}

Here we prove the traceless condition in Eq.~\eqref{eq:tracelessness}. We start from the quadratic Casimir operator $S^\alpha_\beta S^\beta_\alpha$, which is proportional to the identity over any irrep, but the constant factor depends on the irrep itself, therefore offering a way to distinguish different irreps. Writing the quadratic Casimir in terms of the Mathur-Sen bosons we get
\begin{equation}
\begin{split}
S^\alpha_\beta S^\beta_\alpha &= \big (a^{\alpha \dagger} a_\beta - b_\beta^\dagger b^\alpha\big) \big (a^{\beta \dagger} a_\alpha - b_\alpha^\dagger b^\beta\big)\\
&= a^{\alpha \dagger} a_\beta a^{\beta \dagger} a_\alpha  +b^{\beta \dagger} b_\alpha b^{\alpha \dagger} b_\beta   -a^{\alpha \dagger} a_\beta b^{\alpha \dagger} b_\beta- b_\beta^\dagger b^\alpha  a^{\beta \dagger} a_\alpha  \\
&= (a^{\alpha \dagger} a_\alpha)  (a^{\beta \dagger} a_\beta)  + 2 a^{\alpha \dagger} a_\alpha + (b^{\alpha \dagger} b_\alpha)  (b^{\beta \dagger} b_\beta)  + 2 b^{\alpha \dagger} b_\alpha  -  2( a^{\beta \dagger}b_\beta^\dagger)( a_\alpha  b^\alpha ), 
\end{split}
\end{equation}
where we only used the bosonic commutations relations to obtain the last line. Now if we consider a general state with p $a$ bosons and p $b$ bosons, we find that the quadratic Casimir is equivalent to

\begin{equation}
\begin{split}
S^\alpha_\beta S^\beta_\alpha = 2p^2 +4 p    -  2( a^{\beta \dagger}b_\beta^\dagger)( a_\alpha  b^\alpha ).
\end{split}
\end{equation}
Since a  p $a$ boson and p $b$ boson state in general can be a mixture of various $[p',p']$ irreps ($p' < p$), the above expression is not proportional to the identity in general. However, every state  $\ket{\psi_{[p,p]}}$ in the $[p,p]$ irrep should be an eigenstate of  the quadratic Casimir with eigenvalue $2p^2 +4 p$,  \footnote{ see Eq.~(4.87) in Ref.\ \onlinecite{Pfeifer2003su3}, for example, but note that we have a factor 2 difference to their convention. One can also find this value for the quadratic Casimir  by checking what  it gives for the $(a^{1\dagger})^p (b_2^\dagger)^p \ket{0}$ state, which is clearly in the $[p,p]$ irrep.   }
therefore
\begin{equation}
\begin{split}
S^\alpha_\beta S^\beta_\alpha \ket{\psi_{[p,p]}} =(2p^2 +4 p) \ket{\psi_{[p,p]}}  -  2( a^{\beta \dagger}b_\beta^\dagger)( a_\alpha  b^\alpha )\ket{\psi_{[p,p]}} = (2p^2 +4 p) \ket{\psi_{[p,p]}} \end{split}.
\end{equation}
From this it straightforwardly follows that  $-  2( a^{\beta \dagger}b_\beta^\dagger)( a_\alpha  b^\alpha )\ket{\psi_{[p,p]}} = 0$ must be fulfilled. This is only true if  $( a_\alpha  b^\alpha )\ket{\psi_{[p,p]}} = 0$, since if we get some  nonzero state after applying the annihilation operators, acting with the creation operators on top of that will also give some nonzero state.

To show that the spin-coherent states defined in Eq.~\eqref{eq:coherentstate} are indeed in the $[p,p]$ irrep,  we need to check if they vanish under $a_\gamma b^\gamma$.

\begin{equation}
\begin{split}
\sum_\gamma a_\gamma b^\gamma  |\vec z,\vec w\rangle =\sum_\gamma a_\gamma b^\gamma    [(\vec z\cdot \vec a^\dagger )(\vec w\cdot \vec b^\dagger )]^p|0\rangle  = p^2( \vec z \cdot \vec w)  [(\vec z\cdot \vec a^\dagger )(\vec w\cdot \vec b^\dagger )]^{p-1}|0\rangle,
\end{split}
\end{equation}
where the right hand side vanishes  when  $\vec z \cdot \vec w=0$.

\section{Detailed description of the flavour-wave modes} 
\label{app:modes}
Each of the six Goldstone modes $\omega_{1,\dots,6}$ of Eq.~\eqref{eq:su3-hw-w-1D-self-conjugated} can  be associated with one of the six off-diagonal generators acting on
the initial condensate. For instance, the modes $\omega_{5,6}$ stemming from the matrix $M^{2}$ arise from the Holstein-Primakoff bosons
$a^{3\dagger}(k, \La)$ and $b_3^{\dagger}(k, \Lb)$. These bosons correspond to acting on the initial condensate
$\frac{1}{p!}\big( a^{1 \dagger}(i) b^{ \dagger}_{2}(i) \big)^{p} \ket{0}_i =: \ket{(A\bar
B)^{\otimes p}}_i$  on sublattice $\La$ with the generator $S^3_1(i)$  or acting on $
\frac{1}{p!}\big( a^{2 \dagger}(j) b^{\dagger}_{1}(j) \big)^{p} \ket{0}_j =: \ket{(B\bar A)^{\otimes p}}_j$   on sublattice $\Lb$ with
generator $S^1_3 (j)$, 
respectively. % This yields another state of the irrep which differs by one color.
Similarly, the modes $\omega_{7,8}$ come from bosons $b_3^{\dagger}(k,\La)$ and $a^{3\dagger}(k, \Lb)$ that correspond to acting
with generators $S^2_3(i)$ on sublattice $\La$ and $S ^3_2(j)$ on sublattice $\Lb$ on the initial condensate,
respectively.

The case of bosons $a^{2\dagger}(k, \La), b_1^{\dagger}(k ,\La), a^{1\dagger}(k, \Lb), b_2^{\dagger}(k,\Lb)$ is a bit different.  The
generators $S ^2_1(i)$ and $S ^1_2(j)$ applied on the initial condensates $\ket{(A\bar{B})^{\otimes p}}_i$ and $\ket{(B\bar{A})^{\otimes p}}_j$ 
create the states $\ket{(A\bar{A}-B\bar{B})(A\bar{B})^{\otimes (p-1)}}_i$  and  $\ket{(A\bar{A}-B\bar{B})(B\bar{A})^{\otimes (p-1)}}_j$, respectively. % Dropping the wave function of the condensed bosons, the state $\ket{A\bar A-B\bar B}$ belongs
They belong to a two-dimensional subspace in the weight diagram, as shown in Fig.~\ref{fig:weight-diag} for the case $p=1$ as an example. In terms of the
bosonic operators in the flavour-wave approach these states correspond to $ f^{1\dagger}(i)= \big( a^{2\dagger} (i)- b_1^{\dagger}(i)\big)/ \sqrt{2}$ on
sublattice $\La$ and $ f^{2\dagger}(j) =\big( a^{1\dagger}(j) -b_2^{\dagger}(j)\big)/\sqrt{2}$ on sublattice $\Lb$, which
 give the remaining two propagating Goldstone modes $\omega_{1,2}$ that come from the matrix $M^{1}$. These modes
have a velocity two times larger than the others, due to the fact that the states created by $S ^2_1(i)$ and
$S ^1_2(j)$ from the initial condensates have a norm twice as large as the states created by the other generators discussed above. Alternatively, in
terms of the Holstein-Primakoff bosons, this can be seen in the fact that the operators $S^3_1(i)$, $S^2_3(i)$ ($S^1_3(j)$, $S^3_2(j)$)
 correspond to creation operators $a^{3\dagger}(i)$ and $b_3^{\dagger}(i)$ on sublattice $\La$
($a^{3\dagger}(j)$ and $b_3^{\dagger}(j)$ on sublattice $\Lb)$, but $S^2_1(i)$ ($S^1_2(j)$) corresponds to
$ \sqrt{2} f^{1\dagger}(i)$ on sublattice $\La$ ($\sqrt{2}f^{2\dagger}(j)$ on sublattice $\Lb$).

Finally, the $\omega_{7,8}$ modes are related to the bosonic operators
$g^{1\dagger}(i)= (a^{2\dagger}(i) + b_1^{\dagger}(i))/\sqrt{2}$ on sublattice $\La$ and
$ g^{2\dagger}(j)=(a^{1\dagger} (j)+ b_2^{\dagger}(j))/\sqrt{2}$ on sublattice $\Lb$.  
In harmonic order, applying these on the initial condensates leads to the states $\ket{(A\bar{A}+B\bar{B})(A\bar{B})^{\otimes (p-1)}}_i$ on sublattice $\La$ and
$\ket{(A\bar{A}+B\bar{B})(B\bar{A})^{\otimes (p-1)}}_j$. These states are actually not fully in the p box self-conjugate irrep.  For example, for $p=1$ the $\ket {A\bar A +B\bar B}$ state is a combination of the $\ket{A\bar{A}+B\bar{B}-2C\bar{C}}$ state in the self-conjugate irrep and the  $\ket{A\bar{A}+B\bar{B}+C\bar{C}}$ state in the  singlet irrep. These two states cannot be distinguished in the harmonic order; only higher order corrections can reveal the true nature of the $g$ bosons. Nonetheless, we can understand the flat nature of these modes by realizing  that $g$ bosons don't correspond to any single generator applied  on the initial condensate. % and $ \ket{(A\bar{A}+B\bar{B}-2C\bar{C})(B\bar{A})^{\otimes (p-1)}}_j$ on sublattice $\Lb$ i cannot be accessed from the initial condensates $\ket{(A\bar{B})^{\otimes p}}_i$ or $\ket{(B\bar{A})^{\otimes p}}_j$ by applying a single generator, 
%because the matrix elements are actually 0 despite being adjacent points on
%the weight diagram, therefore a product of generators needs to be applied on our initial state 
% \begin{equation}
%   \label{eq:su3-hw-2nd-order-transition}
%   \begin{aligned}
%     \left[ 2 S^{2}_{3}(i) S^{3}_{1}(i) - S^{2}_{1}(i) \right] \ket{\varphi_{i}} =& \left( a ^{1 \dagger}_{2} a ^{1 }_{1} + b
%       ^{1 \dagger}_{1} b ^{1 }_{2} - 2 b ^{1 \dagger}_{3} b ^{1 }_{2} a ^{1 \dagger}_{c} a ^{1 }_{1}\right)
%     \ket{\varphi_{i}}\\
%     \equiv& (A\bar{A}+B\bar{B}-2C\bar{C})(A\bar{B})^{\otimes (p-1)}
%   \end{aligned}
% \end{equation}
For example the $\ket{A\bar{A}+B\bar{B}-2C\bar{C}}$ state can be reached from $\ket{ A\bar B}$ by applying   $\frac{2}{\sqrt{6}} S^{2}_{3}(i) S^{3}_{1}(i) - \frac{1}{\sqrt{6}}S^{2}_{1}(i)$ ($S^{2}_{1}$ itself only leads to $\ket{A\bar A- B\bar B}$). 
Such multipolar states requiring a product of generators on one site are well known in the literature.\cite{Romhanyi2012,Penc2012}   They appear as flat modes in the harmonic order, because dispersive terms  appear in the expansion of the off-diagonal
interaction terms only in higher order. For the same reason, the $\omega_{7}$ and $\omega_{8}$ stemming from the matrix $M^{1}$ are flat, and they do not correspond to the Goldstone modes.

Formally rewriting the Hamiltonian  in Eq.~\eqref{eq:su3-hw-H2} using the new bosons $f,g$ bosons
\begin{equation}
  \begin{split}
    \label{eq:su3-hw-f}
 f_1(i)&= \frac{1}{\sqrt{2}}\big( a_{2} (i)- b^1(i)\big),\\
    f_2(j)&=\frac{1}{\sqrt{2}} \big( a_{1}(j) - b^2(j) \big),\\
    g_1(i)&=\frac{1}{\sqrt{2}} \big( a_{2}(i) + b^1(i) \big),\\
    g_2(j)&=\frac{1}{\sqrt{2}} \big( a_{1}(j) + b^2(j) \big),\\
  \end{split}
\end{equation}
we find
\begin{equation}
  \label{eq:su3-hw-H2-f}
  \begin{aligned}
    \mathcal{H}^{(2)} = J n_c p \sum\limits_{i \in \Lambda_{1}}\sum\limits_{\left\langle j \right\rangle} &\Big\{
      2 g ^{1\dagger}(i) g _{1}(i) + 2 g^{2\dagger}(j) g _{2}(j)  \\
      &+ 2 \Big[ f ^{1\dagger}(i) f _{1}(i) + f ^{2\dagger}(j) f _{2}(j) + f ^{1\dagger}(i) f ^{2\dagger}(j) + f _1(i) f _2(j)\Big] \\
      &+ \left[  a ^{3\dagger}(j) a ^{}_{3}(j) +  b ^{\dagger}_{3}(i) b ^{3}(i) -  b ^{\dagger}_{3}(i) a
        ^{3\dagger}(j) - b ^{3}(i) a_{3}(j) \right]\\
      &+ \left[ a ^{3\dagger}(i) a_{3}(i) + b^{\dagger}_{3}(j) b ^{3}(j) - a^{3\dagger}(i) b
          ^{\dagger}_{3}(j) - a_{3}(i) b ^{3}(j) \right] \Big\},
  \end{aligned}
\end{equation}
where the new $g_1, g_2$ bosons clearly give non-dispersive flat modes. %These modes will gain dispersion if we include  higher order terms in the spin wave expansion. 
To satisfyingly study the true nature of these modes, i.e. if they truly belong to the self-conjugate irrep or not requires higher order spin wave corrections, but it doesn't change our conclusions, since we focus on the low energy modes.
  
  \section{Detailed calculations for the  derivation of the NLI$\sigma$M}
  \label{app:SLU}
  
  Here we show how different terms read in the expansion of the Hamiltonian and Berry phase terms.  The $\vec \phi$ fields in the $j+1$th unit cell are expanded as $\vec \phi^n_{j+1} = \vec\phi^n_j+2a \partial_x \vec \phi^n_j + 2a^2  \partial_x^2 \vec \phi^n_j+\mathcal{O}(a^3)$. The $\tau$ imaginary time variable  and the $j$ unit cell index on the right hand side are omitted for simplicity.
\begin{equation}
\begin{split}
p^2|\vec \Phi^{1*}_{2j} \cdot\vec \Phi^2_{2j+1}|^2 &=4 a^2 |L_{12}|^2+\mathcal{O}(a^3/p)  \\
p^2 |\vec \Phi^{2*}_{2j}\cdot \vec \Phi^1_{2j+1}|^2 &=4 a^2 |L_{12}|^2 +\mathcal{O}(a^3/p)  \\
- p^2|\vec \Phi^{1*}_{2j} \cdot\vec \Phi^1_{2j+1}|^2 &=-p^2 +4a^2 |L_{12}|^2+ 4a^2 |L_{13}|^2+\mathcal{O}(a^4/p^2)\\
- p^2|\vec \Phi^{2*}_{2j} \cdot\vec \Phi^2_{2j+1}|^2 &= -p^2 + 4a^2 |L_{12}|^2+ 4a^2 |L_{23}|^2 +\mathcal{O}(a^4/p^2)
\end{split}
\end{equation}

\begin{equation}
\begin{split}
p^2\vec |\Phi^{2*}_{2j+1} \cdot \vec \Phi^1_{2j+2}|^2
&=|2aL_{12} + 2ap (\vec \phi^{2*} \cdot\partial_x \vec \phi^1) + \mathcal{O}(a^2)|^2 \\
&=4a^2|L_{12}|^2+4a^2p^2|\vec \phi^{2*} \cdot\partial_x \vec \phi^1|^2
+4a^2pL^*_{12}(\vec \phi^{2*} \cdot\partial_x \vec \phi^1)
+4a^2pL_{12}(\vec \phi^2 \cdot\partial_x \vec \phi^{1*})+\mathcal{O}(a^3) \\
&= p^2\vec |\Phi^{1*}_{2j+1} \cdot \vec \Phi^2_{2j+2}|^2
\end{split}
\end{equation}

where we used 
\be \vec \phi^{1*}\cdot \partial_x\vec \phi^2+\vec \phi^2\cdot \partial_x\vec \phi^{1*}=\partial_x(\vec \phi^{1*}\cdot \vec \phi^2)=0.\ee
\begin{equation}
\begin{split}
\vec \Phi^{2}_{2j+1} \cdot\vec \Phi^{2*}_{2j+2}=&{}-{a^2\over p^2}|L_{12}|^2 -{a^2\over p^2} |L_{23}|^2
+ \big(1-{a^2\over p^2}|L_{12}|^2 -{a^2\over p^2}|L_{23}|^2\big)\big(1 + 2a (\vec \phi^2\cdot\partial_x \vec \phi^{2*})
+2a^2 (\vec \phi^2\cdot\partial_x^2 \vec \phi^{2*}) \big)\\
&+\frac{2a^2}{p} L_{12}^* (\vec \phi^1 \cdot\partial_x \vec \phi^{2*})
-\frac{2a^2}{p} L_{12} (\vec \phi^2\cdot\partial_x 
\vec \phi^{1*}) -\frac{2a^2}{p}  L_{23} (\vec \phi^3\cdot\partial_x \vec \phi^{2*}) 
+\frac{2a^2}{p}  L_{23}^* (\vec \phi^2\cdot\partial_x \vec \phi^{3*}) \\
&+\mathcal{O}(a^3).
\end{split}
\end{equation}
\begin{equation}
\begin{split}
 -p^2|\vec \Phi^2_{2j+1}\cdot \vec \Phi^{2*}_{2j+2}|^2=&{} 
-p^2+4a^2 |L_{12}|^2 +4a^2 |L_{23}| ^2 -4a^2p^2  |\vec \phi^{2*} \cdot\partial_x \vec \phi^2|^2 \\
&-2a^2 p^2\big( \vec \phi^2 \cdot\partial_x^2 \vec \phi^{2*}+\vec \phi^{2*} \cdot\partial_x^2 \vec \phi^2\big)
 -2a p^2 (\vec \phi^2 \cdot\partial_x \vec \phi^{2*}+\vec \phi^{2*} \cdot\partial_x \vec \phi^2 )\\
&-4a^2 p L_{12}^*(\vec \phi^1 \cdot\partial_x \vec \phi^{2*}) +4a^2 p L_{23} (\vec \phi^3 \cdot\partial_x \vec \phi^{2*})
\nonumber \\
&-4a^2 p L_{12} (\vec \phi^{1*} \cdot\partial_x \vec \phi^2) +4a^2 p L_{23}^* (\vec \phi^{3*} \cdot\partial_x \vec \phi^2)
+\mathcal{O}(a^3) \nonumber \\
=&{}-p^2+4a^2 |L_{12}|^2 +4a^2 |L_{23}| ^2 +4a^2p^2 ( |\partial_x \vec \phi^2|^2- |\vec \phi^{2*} \cdot\partial_x \vec \phi^2|^2)  \\
&+ 4a^2 p L_{12} (\vec \phi^2 \cdot\partial_x \vec \phi^{1*}) -4a^2 p L_{23} (\vec \phi^{2*} \cdot\partial_x \vec \phi^3)
\nonumber \\
&+4a^2 p L_{12}^* (\vec \phi^{2*} \cdot\partial_x \vec \phi^1) -4a^2 p L_{23} ^*(\vec \phi^2 
\cdot\partial_x \vec \phi^{3*}) +\mathcal{O}(a^3),
\end{split}
\end{equation}
where we used 
\be
\partial^2_x \vec \phi^{n*}\cdot \vec \phi^n+2 \partial_x \vec \phi^{n*}\cdot \partial_x\vec \phi^n+ \vec \phi^{n*} \cdot \partial_x^2 \vec \phi^{n}= \partial_x^2( |\vec \phi^n|^2)=0.
\ee
Finally
\begin{equation}
\begin{split}
\vec \Phi^{1*}_{2j+1} \cdot\vec \Phi^1_{2j+2}=&{}-{a^2\over p^2}|L_{12}|^2-{a^2\over p^2} |L_{13}|^2+ 
\big(1-{a^2\over p^2} (|L_{12}|^2+|L_{13}|^2)\big)\big(1 + 2a (\vec \phi^{1*} \cdot\partial_x \vec \phi^1)+
2a^2 (\vec \phi^{1*} \cdot\partial_x^2 \vec \phi^1) \big) \\
&+\frac{2a^2}{p} L_{12} (\vec \phi^{1*} \cdot\partial_x \vec \phi^2)-\frac{2a^2}{p} L_{12}^* (\vec \phi^{2*}
 \cdot\partial_x \vec \phi^1) +\frac{2a^2}{p} L_{13} (\vec \phi^{1*}\cdot\partial_x \vec \phi^3)-\frac{2a^2}{p} L_{13}^*
(\vec \phi^{3*}\cdot\partial_x \vec \phi^1) +\mathcal{O}(a^3)
\end{split}
\end{equation}
\begin{equation}
\begin{split}
-p^2|\vec \Phi^{1*}_{2j+1} \cdot\vec \Phi^1_{2j+2}|^2 =&{} -p^2+4a^2 |L_{12}|^2 +4a^2 |L_{13}| ^2   
+4a^2 p^2(|\partial_x \vec \phi^1|^2- |\vec \phi^{1*}\cdot\partial_x \vec \phi^1|^2) \\
&+4a^2 pL_{12} (\vec \phi^2 \cdot\partial_x \vec \phi^{1*}) +4a^2 p L_{13}^* 
(\vec \phi^{3*} \cdot\partial_x \vec \phi^1) \\
&+4a^2 p L_{12}^* (\vec \phi^{2*} \cdot\partial_x \vec \phi^1) +4a^2 p L_{13} (\vec \phi^3 \cdot\partial_x \vec \phi^{1*}) 
+\mathcal{O}(a^3)
\end{split}
\end{equation}

The Berry's phase terms become:
\begin{equation}
\begin{split}
\vec \Phi^{1*}_{2j}\cdot \partial_\tau \vec \Phi^1_{2j}&-\vec \Phi^{1*}_{2j+1}\cdot \partial_\tau \vec \Phi^1_{2j+1}
-\vec \Phi^{2*}_{2j}\cdot \partial_\tau \vec \Phi^2_{2j}+\vec \Phi^{2*}_{2j+1}\cdot \partial_\tau \vec \Phi^2_{2j+1} \\
\approx {2a\over p} \Big(&2L_{12}(\vec \phi^{1*}\cdot \partial_\tau \vec \phi^2)+2L_{12}^*(\vec \phi^{2*}\cdot \partial_\tau \vec \phi^1)
+L_{13}(\vec \phi^{1*}\cdot \partial_\tau \vec \phi^3)\\
&+L_{13}^*(\vec \phi^{3*}\cdot \partial_\tau \vec \phi^1)
-L_{23}(\vec \phi^{2*}\cdot \partial_\tau \vec \phi^3)-L_{23}^*(\vec \phi^{3*}\cdot \partial_\tau \vec \phi^2)\Big).
\end{split}
\end{equation}

Using the above expansions we arrive to the action given in Eq.~\eqref{eq:SLU}. The $L$ fields can be integrated out using the Gaussian identity
\begin{equation}
\begin{split}
\int dzdz^* \exp\big(- z^* \omega z + u^*z+vz^*\big) = \frac{\pi}{\omega} \exp\big(\frac{u^*v}{\omega}\big).
\end{split}
\end{equation}
Carrying out the Gaussian integrals in $L$ fields gives
\begin{equation}
\begin{split}
S=  \int dx d\tau  \Bigg(&a p^2 J\Big[4| \vec \phi^{2*}\cdot\partial_x \vec \phi^1|^2
+2\big(|\partial_x \vec \phi^1|^2-|\vec \phi^{1*}\cdot\partial_x \vec \phi^1|^2 \big)+
2 \big(|\partial_x \vec \phi^2|^2-|\vec \phi^{2*}\cdot\partial_x \vec \phi^2|^2 \big) \\
&-4| \vec \phi^{2*}\cdot\partial_x \vec \phi^1|^2 - 
 | \vec \phi^{2*}\cdot\partial_x \vec \phi^3|^2-| \vec \phi^{1*}\cdot\partial_x \vec \phi^3|^2\Big] \\
&+\frac{1}{4aJ} \Big[| \vec \phi^{2*}\cdot\partial_\tau \vec \phi^1|^2+
| \vec \phi^{2*}\cdot\partial_\tau \vec \phi^3|^2+| \vec \phi^{1*}\cdot\partial_\tau \vec \phi^3|^2\Big] \\
&-p \Big[(\vec \phi^{2*}\partial_x \vec \phi^1)(\vec \phi^{1*}\partial_\tau\vec \phi^2)- 
(\vec \phi^{2*}\partial_\tau \vec \phi^1)(\vec \phi^{1*}\partial_x\vec \phi^2)  \\
&+ \frac{1}{2}(\vec \phi^{2*}\partial_x \vec \phi^3)(\vec \phi^{3*}\partial_\tau\vec \phi^2)- 
\frac{1}{2}(\vec \phi^{2*}\partial_\tau \vec \phi^3)(\vec \phi^{3*}\partial_x\vec \phi^2)  \\
 &+ \frac{1}{2} (\vec \phi^{3*}\partial_x \vec \phi^1)(\vec \phi^{1*}\partial_\tau\vec \phi^3)
- \frac{1}{2}(\vec \phi^{3*}\partial_\tau \vec \phi^1)(\vec \phi^{1*}\partial_x\vec \phi^3) \Big]\Bigg),
\end{split}
\end{equation}

that leads to Eq.~\eqref{eq:symmaction}.

 \section{Order parameters for  spontaneous symmetry breaking}
 \label{app:symmbreak}
 
Here we present various order parameters that could signal the  spontaneous breakdown of the  various discrete symmetries discussed in Sec.~\ref{sec:symmetries}.  We are looking for expressions which should give 0 if a symmetry is conserved, therefore a nonzero value would mean the breakdown of that given symmetry. We suggest two families of order parameters, one based on topological charges, and one based on $\mbox{SU}(3)$ and gauge invariant terms built from the  $\vec \phi$ fields of nearest neighbour sites of the discretized action. 

Take for example the $\mathbb{Z}_2^{(\text{tr})}$ symmetry related to the translation by one site. Since this symmetry maps $Q_1$ to $Q_2$ and vice versa, $\langle Q_1-Q_2\rangle $ changes sign under this transformation, therefore if translational symmetry is conserved this quantity should be 0. On the other hand, if $\langle Q_1 - Q_2\rangle \neq 0$, it would suggest that the translational invariance is explicitly broken.  

Following a similar argument we can construct order parameters from the $\vec \phi$ fields as well. Consider for example the term $ | \vec \phi^{1*}( x,\tau) \cdot \vec \phi^3(x+\delta_x,\tau)|^2$ on a discretized lattice, under  $\mathbb{Z}_2^{(\text{tr})}$ this maps to $ | \vec \phi^{2*}( x,\tau) \cdot \vec \phi^3(x+\delta_x,\tau)|^2$ and vice versa. Therefore  if $\mathbb{Z}_2^{(\text{tr})}$ is conserved,  the expectation $\big\langle 
\sum_{x,\tau}\big( | \vec \phi^{1*}( x,\tau) \cdot \vec \phi^3(x+\delta_x,\tau)|^2 - | \vec \phi^{2*}( x,\tau) \cdot \vec \phi^3(x+\delta_x,\tau)|^2 \big)\big\rangle
$
should vanish, or if it  is nonzero it means that  $\mathbb{Z}_2^{(\text{tr})}$ is broken.  Following the same argument we can find multiple {\it canditate} order parameters for the breakdown of each of the previously discussed symmetries. If any of the candidate order parameters of a given symmetry are nonzero, it follows that the symmetry must be broken.  %However, the breakdown of a given symmetry doesn't mean that any of the related candidate order parameters must be nonzero. 

Based on Griffith,\cite{Griffith1966} we propose measuring the correlations of the local order parameters, and extracting the long distance limit of these correlations from finite size simulations. If this long distance limit  is nonzero  for a given order parameter that suggests spontaneous  breakdown of the associated symmetry in the thermodynamic limit. In Table \ref{tab:OP} we provide a list of candidate local order parameters for each of the symmetries discussed in Sec.~\ref{sec:symmetries}. 

{\renewcommand{\arraystretch}{1.5}
\begin{table}[htp]
\begin{center}
\begin{tabular}{|c|@{\hspace{0.6em}}c@{\hspace{0.6em}} |@{\hspace{1.2em}}l@{\hspace{1.2em}} |}
\hline
symmetry & \multicolumn{2}{c|}{candidate order parameters}\\
\hline
{$\mathbb{Z}_2^{(\text{tr})}$} & {$  q_1(\vec r)-q_2( \vec r)  $} &  $ \oa_1^{x(\tau)}(\vec r)=   | \vec \phi^{1*}( \vec r) \cdot \vec \phi^3(\vec r  +\vec \delta_{x(\tau)})|^2 - | \vec \phi^{2*}( \vec r) \cdot \vec \phi^3(\vec r +\vec \delta_{x(\tau)})|^2 $\\
&&  $ \oa_2^{x(\tau)}(\vec r)=   | \vec \phi^{1*}( \vec r) \cdot \vec \phi^2(\vec r  +\vec \delta_{x(\tau)})|^2 - | \vec \phi^{2*}( \vec r) \cdot \vec \phi^1(\vec r +\vec \delta_{x(\tau)})|^2 $\\
 && $ \oa_3^{x(\tau)}(\vec r)=   | \vec \phi^{1*}( \vec r) \cdot \vec \phi^1(\vec r  +\vec \delta_{x(\tau)})|^2 - | \vec \phi^{2*}( \vec r) \cdot \vec \phi^2(\vec r +\vec \delta_{x(\tau)})|^2 $\\
\hline
{$\mathbb{Z}_2^{(\text{sp})}$} & $   q_1(\vec r)   $, $   q_2(\vec r)   $, $   q_3(\vec r)   $ &$ \ob_1^x(\vec r)=  | \vec \phi^{1*}( \vec r) \cdot \vec \phi^3(\vec r  +\vec \delta_x)|^2 - | \vec \phi^{1*}( \vec r+\vec \delta_x) \cdot \vec \phi^3(\vec r)|^2 $\\
& &$ \ob^x_2(\vec r)=  | \vec \phi^{2*}( \vec r) \cdot \vec \phi^3(\vec r  +\vec \delta_x)|^2 - | \vec \phi^{2*}( \vec r+\vec \delta_x) \cdot \vec \phi^3(\vec r)|^2 $\\
%&& $\oa_2^{x}(\vec r)$\\
&&$ \ob^x_3(\vec r)=  | \vec \phi^{1*}( \vec r) \cdot \vec \phi^2(\vec r  +\vec \delta_x)|^2 - | \vec \phi^{1*}( \vec r+\vec \delta_x) \cdot \vec \phi^2(\vec r)|^2 \equiv \oa_2^{x}(\vec r)$\\
\hline
{$\mathbb{Z}_2^{(\text{bp})}$} & $   q_1(\vec r)  +q_2(\vec r) $ 
&$ \oc_1^x(\vec r)=  | \vec \phi^{1*}( \vec r) \cdot \vec \phi^3(\vec r  +\vec \delta_x)|^2 - | \vec \phi^{2*}( \vec r+\vec \delta_x) \cdot \vec \phi^3(\vec r)|^2 $\\
& &$ \oc^x_2(\vec r)=  | \vec \phi^{1*}( \vec r) \cdot \vec \phi^2(\vec r  +\vec \delta_x)|^2 - | \vec \phi^{2*}( \vec r+\vec \delta_x) \cdot \vec \phi^1(\vec r)|^2 \equiv 0$\\
& &$ \oc^x_3(\vec r)=  | \vec \phi^{1*}( \vec r) \cdot \vec \phi^1(\vec r  +\vec \delta_x)|^2 - | \vec \phi^{2*}( \vec r+\vec \delta_x) \cdot \vec \phi^2(\vec r)|^2 \equiv \oa_3^x(\vec r)$\\
&&  $\oa_1^{\tau}(\vec r)$, $\oa_2^{\tau}(\vec r)$, $\oa_3^{\tau}(\vec r)$ \\
\hline
{$\mathbb{Z}_2^{(a\leftrightarrow b)}$} & $  q_1(\vec r)  +q_2(\vec r) $ 
& $\oa_1^{x(\tau)}(\vec r)$, $\oa_2^{x(\tau)}(\vec r)$, $\oa_3^{x(\tau)}(\vec r)$ \\
\hline
\end{tabular}
\caption{A list of candidate order parameters for detecting the breakdown of various  discrete symmetries. From top to bottom these are the $\mathbb{Z}_2$ symmetries related to translation (tr), site parity (sp), bond parity (bp), and the $a_\alpha\leftrightarrow b^\alpha$ parity.    $q_i(\vec r)$ stands for the local topological charge density on a plaquette, where the total topological charge is $Q_i = \sum_{\vec r} q_i(\vec r)$. The superscript of the field-based order parameters shows whether they are defined on bonds in the spatial or imaginary time direction. Note that some order parameters appear for multiple symmetries,  therefore if they are measured to be nonzero, it would  suggest the breakdown of all related  symmetries. \label{tab:OP}}
\end{center}
\end{table}
}

For the current model the most relevant  candidates are $ \big\langle q_1(\vec r)  +q_2(\vec r) \big \rangle$  and $\oc_1^x(\vec r)$. As we discussed above we believe that the {$\mathbb{Z}_2^{(\text{bp})}$} bond parity and  {$\mathbb{Z}_2^{(a\leftrightarrow b)}$} symmetries are spontaneously broken for $\theta_1=-\theta_2=\pi$. This is  supported by the strong coupling calculations in Sec.~\ref{sec:strongcoupling}, where we showed that  $\langle q_1(\vec r) +q_2(\vec r)\rangle$ is nonzero in the thermodynamic limit, and  by  the AKLT-type example of Sec.~\ref{sec:AKLT}.  However, both the strong coupling calculations and the AKLT examples show that the $\mathbb{Z}_2^{(\text{tr})}$ remains conserved, which would fix all $\oa$ order parameters to 0. 
Unfortunately, so far we haven't been able to obtain clear results on the breakdown of symmetries for the physically relevant $\theta_1=-\theta_2=\pi$ in the weak coupling case  using the extrapolation technique of Sec.~\ref{sec:MC}. We believe that the  main obstacle lies in the extrapolation itself. For  imaginary topological angles we find that the correlations of the local order parameters converge to a fixed value  within a few lattice spacing, promising a good estimate on the infinite range correlation. We found that the  long range correlations of all order parameters go to 0 at $\theta_1=-\theta_2=0$, but we couldn't get reliable estimates for finite real angles. We needed higher degree polynomials to accurately fit the results for imaginary angles, but as a result the extrapolated values  were really sensitive to the fitting parameters, which made it impossible to get reliable values. 

\bibliographystyle{apsrev4-1}
\bibliography{su3qftSC.bib}

\end{document}